\documentclass[12pt,a4paper,oneside]{extarticle}

\usepackage{cmap}	
\usepackage{graphicx}
\usepackage{subfiles}
\usepackage{amsthm}
\usepackage{amsmath}
\allowdisplaybreaks
\usepackage{amssymb}
\usepackage{xcolor}
\usepackage{booktabs}
\usepackage{array}
\usepackage{caption}
\usepackage[T2A]{fontenc}
\usepackage{amsfonts}
\usepackage[breakable]{tcolorbox}
\tcbuselibrary{theorems}
\usepackage[english]{babel}
\usepackage{listings}

\usepackage{geometry}%The package provides an easy and flexible user interface to customize page layout, implementing auto-centering and auto-balancing mechanisms so that the users have only to give the least description for the page layout.
\usepackage[numbers,square,sort]{natbib}
\bibpunct{(}{)}{,}{a}{,}{,}
\usepackage[colorlinks=true, linkcolor=black, citecolor=black, urlcolor=blue]{hyperref}

\lstdefinestyle{rstudio}{
  language=R,
  basicstyle=\ttfamily\small,
  commentstyle=\color{gray}\itshape,
  keywordstyle=\color{black},
  stringstyle=\color{green!40!black},
  numberstyle=\color{purple},
  identifierstyle=\color{black},
  backgroundcolor=\color{white},
  showstringspaces=false,
  frame=none,
  breaklines=true,
  morekeywords={library,summary,lm,glm,plot},
}

\lstnewenvironment{Sinput}
  {\lstset{style=rstudio}}
  {}

\lstnewenvironment{Soutput}
  {\lstset{basicstyle=\ttfamily\small, language={}}}
  {}

\newcommand{\code}[1]{\texttt{\detokenize{#1}}}

\newcounter{theorem}[subsection]
\renewcommand{\thetheorem}{\thesubsection.\arabic{theorem}}

\newcounter{lemma}[subsection]
\renewcommand{\thelemma}{\thesubsection.\arabic{lemma}}

\newcounter{assumption}[subsection]
\renewcommand{\theassumption}{\thesubsection.\arabic{assumption}}

\newcounter{definition}[subsection]
\renewcommand{\theassumption}{\thesubsection.\arabic{definition}}

\newenvironment{lemma}[1][]
{
    \refstepcounter{lemma}
    \if\relax\detokenize{#1}\relax
        \label{lemma:\thesubsection.\arabic{lemma}}%
    \else
        \label{#1}%
    \fi
    \begin{tcolorbox}[
        breakable,
        colback=white,
        colframe=black,
        boxrule=0.8pt,
        left=10pt,
        right=10pt,
        top=10pt,
        bottom=10pt,
        before skip=12pt,
        after skip=12pt
    ]
    \textbf{Lemma \thelemma.}\\[6pt]
}
{
    \end{tcolorbox}
}

\newenvironment{proofof}
{
    \par\vspace{12pt}
    \noindent
    \ignorespaces
}
{
    \par\nobreak\hfill$\square$\par
    \vspace{12pt}
}

\newcommand{\startproof}[2]{%
    \par\vspace{12pt}%
    \noindent\textbf{Proof #1 \ref{#2}.}\par\medskip%
}

\title{mnorm: An R Package for Calculation and Differentiation of Conditional Multivariate Normal Densities and Probabilities}
\author{Sofiia Dolgikh \\ HSE University \\ 
        \and Bodan Potanin \\ HSE University}
\date{}

\begin{document}

\maketitle

\begin{abstract}
We introduce the \textbf{mnorm} package, which allows one to calculate conditional multivariate normal densities and probabilities and to differentiate them with respect to various parameters including covariances and integration limits. The package also supports parallel (multi-core) computing, handles non-normal marginals via the Gaussian copula, and provides fast routines for the calculation of bivariate and trivariate normal probabilities. The package is of special interest for the implementation of the maximum-likelihood estimators in econometric models.
\end{abstract}


\smallskip


\section{Introduction}

The multivariate normal distribution (MVN) has various applications including maximum-likelihood estimation and Bayesian inference. There are many \textbf{R} packages dealing with MVN. The most popular one is \textbf{mvtnorm} \citep{mvtnorm}. It provides tools for calculation of multivariate normal probabilities and densities. For many practical applications, these tools are sufficient. However, sometimes it is necessary to get a boost in terms of speed, accuracy, or functionality. To increase calculation speed in the low-dimensional case, the \textbf{pbivnorm} package may be used, which provides fast calculation of bivariate normal probabilities. The \textbf{mvnfast} package allows for fast parallel (multi-core) calculation of MVN densities. To calculate conditional or truncated MVN probabilities, one may use the \textbf{condMVNorm} and \textbf{TruncatedNormal} packages, respectively. The \textbf{tlrmvnmvt} package \citep{tlrmvnmvt} implements methods suitable for accurate calculation of MVN probabilities in the high-dimensional case. To analyse MVN data, one may use the \textbf{norm} and \textbf{MVN} \citep{MVN} packages. The \textbf{copula} package \citep{copula} allows one to simulate from the Gaussian copula, which extends MVN to the case of non-normal margins.

The \textbf{mnorm} package aims to extend the MVN tools available in R mainly by providing functions for differentiation (calculation of gradients) of MVN probabilities and densities with respect to various parameters including covariances and integration limits. Particularly, it may be useful for maximization of likelihood functions for which gradient information may substantially speed up a numeric optimization routine. For example, the \textbf{switchSelection} package uses \textbf{mnorm} for estimation of multivariate and multinomial sample selection and endogenous switching econometric models.

Also, the package allows one to calculate and differentiate conditional MVN probabilities. Particularly, it is possible to calculate gradients with respect to conditioned (given) components of the distribution (elements of the MVN vector). Moreover, a user may provide an alternative (non-normal) specification of marginal distributions (i.e., deal with the Gaussian copula) and differentiate probabilities with respect to the parameters of these distributions. To facilitate speed, the package is written in \textbf{Rcpp} \citep{Rcpp} and intensively utilizes \textbf{RcppArmadillo} \citep{RcppArmadillo}. It also uses Open Multi-Processing (OMP) for parallel computing.

The article proceeds as follows. In Section \ref{sec:mvn}, we provide theoretical information on the MVN distribution. In Section \ref{sec:examples}, we demonstrate the application of \code{dmnorm} and \code{pmnorm} functions allowing one to calculate and differentiate (conditional) MVN densities and probabilities, respectively. In Section \ref{sec:efficiency}, we compare the efficiency (speed) of our analytical gradients to the lower bound time needed to compute numeric gradients. Moreover, we compare the calculation speed of densities and probabilities to that of other popular packages. In Section \ref{sec:derivatives}, we provide the derivation of the differentiation formulas implemented in the package. In Section \ref{sec:bounds}, we derive some simple, useful bounds on the MVN log-probabilities and their derivatives, which may accompany the usage of the package. Particularly, we contribute to the literature on the bounds for inverse Mills ratios \citep{From2020} by providing a useful bound on the derivatives of log-probabilities.

\section{Multivariate normal distribution}\label{sec:mvn}

Consider an $m$-variate random row vector which has the multivariate normal distribution:
\begin{equation}\label{eq:1}
X = \left(X_{1},...,X_{m}\right)\sim\text{N}\left(\mu,\Sigma\right).
\end{equation}

Parameters of this distribution are the row vector of expectations and the covariance matrix:
\begin{align}
\mu &= \left(\mu_{1},...,\mu_{m}\right)=\mathbb{E}(X)=\left(\mathbb{E}(X_{1}),...,\mathbb{E}(X_{m})\right),\label{(eq:2)}\\
\Sigma &= \text{Cov}(X)=
\begin{bmatrix}
\text{Var}(X_{1}) & \text{Cov}(X_{1},X_{2}) & ... & \text{Cov}(X_{1},X_{m})\\
\text{Cov}(X_{2},X_{1}) & \text{Var}(X_{2}) & ... & \text{Cov}(X_{2},X_{m})\\
\vdots      & ...         & ... & ...        \\
\text{Cov}(X_{m},X_{1}) & \text{Cov}(X_{m},X_{2}) & ... & \text{Var}(X_{m})
\end{bmatrix}\label{(eq:3)}.
\end{align}

The density function of the multivariate normal vector $X$ is as follows:
\begin{equation}
f(x;\mu,\Sigma) = \frac{1}{\sqrt{\left(2\pi\right)^{m}\text{det}\left(\Sigma\right)}}
\exp\left(-\frac{1}{2}\left(x-\mu\right)^T \Sigma \left(x-\mu\right)\right),\label{eq:4}
\end{equation}
where $x = (x_{1},...,x_{m})$ is a row vector and $\text{det}\left(\Sigma\right)$ is the determinant of the covariance matrix $\Sigma$.

Probabilities of the multivariate normal distribution are as follows:
\begin{align}
\overline{F}\left(x^{(l)}, x^{(u)};\mu,\Sigma\right) &= \mathbb{P}\left(x^{(l)}\leq X\leq x^{(u)}\right) = \nonumber \\
&= \mathbb{P}\left(x^{(l)}_{1}\leq X_{1}\leq x^{(u)}_{1},...,x^{(l)}_{m}\leq X_{m}\leq x^{(u)}_{m}\right) = \nonumber \\
&=\int\limits_{x^{(l)}_{1}}^{x^{(u)}_{1}}...\int\limits_{x^{(l)}_{m}}^{x^{(u)}_{m}} f(x;\mu, \Sigma) dx_{1}...dx_{m},\label{eq:5}
\end{align}
where $x^{(l)}_{j}, x^{(u)}_{j}\in \mathbb{R}\cup\{-\infty,\infty\}$ are the elements of the row vectors $x^{(l)}$, $x^{(u)}$, representing lower and upper integration limits, respectively.

Suppose that some of the components of $X$ are conditioned. Denote by $I_{g}$ a vector of indices of the conditioned components. The following vector has a conditional multivariate normal distribution:
\begin{equation}
    \left(X_{d} | X_{g} = x^{(g)}\right) \sim \text{N}\left(\mu_{c},\Sigma_{c}\right),\label{eq:6}
\end{equation}
where the vector of conditional expectations and the conditional covariance matrix are as follows:
\begin{align}
\mu_{c} &= \mathbb{E}\left(X_{d} | X_{g} = x^{(g)}\right) = 
                        \mu_{d} + 
                        \left(x^{(g)} - \mu_{g}\right)
                        \left(\Sigma_{(d, g)}
                              \Sigma_{(g, g)}^{-1}\right)^{T},\label{eq:7}\\
\Sigma_{c} &= \text{Cov}\left(X_{d} | X_{g} = x^{(g)}\right) = 
                             \Sigma_{(d, d)} - 
                             \Sigma_{(d, g)}
                             \Sigma_{(g, g)}^{-1}
                             \Sigma_{(g, d)}\label{eq:8}.
\end{align}

Denote by $\Sigma_{(A, B)}$ a submatrix of $\Sigma$, where $A,B\in\{d, g\}$. This matrix is constructed from $\Sigma$ by the intersection of rows and columns with indices $I_{A}$ and $I_{B}$, respectively. Particularly, $\Sigma_{d,g}^{T}=\Sigma_{g,d}$.

Random vectors $X_{d}$ and $X_{g}$ are subvectors of $X$, with elements with indices $I_{d}$ and $I_{g}$, respectively. Similarly, $\mu_{d}$ and $\mu_{g}$ are subvectors of $\mu$. The density function of the conditional multivariate normal distribution at point $x^{(d)}$ is denoted by $f\left(x^{(d)};\mu,\Sigma | x^{(g)},I_{g}\right)$, where the row vector $x^{(g)}$ contains elements at which the elements of $X_{g}$ are conditioned. 

The numbers of conditioned and unconditioned elements of $X$ are as follows:
\begin{align}
    m_{d}=\sum\limits_{k=1}^{m}\mathbb{I}(k\in I_{d}), \\ 
    m_{g}=\sum\limits_{k=1}^{m}\mathbb{I}(k\in I_{g}).
\end{align}

Consider $i\in I_{d}$ and $j\in I_{g}$. Also, denote by $I_{g,k}$ and $I_{d,k}$ the $k$-th elements of the vectors $I_{g}$ and $I_{d}$, respectively. Let
\begin{align}
    q_{d}(i)&=\sum\limits_{k=1}^{m_{d}} \mathbb{I}\left(I_{d,k} \leq i\right), \\
    q_{g}(j)&=\sum\limits_{k=1}^{m_{g}} \mathbb{I}\left(I_{g,k} \leq j\right),
\end{align}
where $q_{d}(i)$ is the index of the $i$-th element of $X$ in $I_{d}$, and $q_{g}(j)$ is the index of the $j$-th element of $X$ in $I_{g}$; hence $x_{i}=x^{(d)}_{q_{d}(i)}$ and $x_{j}=x^{(g)}_{q_{g}(j)}$. Finally, denote by $q_{g}^{-1}(\cdot)$ and $q_{d}^{-1}(\cdot)$ the inverse functions of $q_{g}(\cdot)$ and $q_{d}(\cdot)$, respectively. For example, let $I_{d} = \{1,3,5\}$ and $I_{g} = \{2, 4\}$. Then $q_{d}(3)=2$ and $q_{g}(2) = 1$. Moreover, $q_{d}^{-1}(2)=3$ and $q_{g}^{-1}(1) = 2$.

The notation $M(a, b; m_{1}, m_{2})$ corresponds to an $\left(m_{1}\times m_{2}\right)$ matrix whose elements are $0$, except for the $(a, b)$-th element, which is equal to $1$. Similarly, $M(a, b; m_{1})$ is a square matrix $\left(m_{1}\times m_{1}\right)$ of zeros, except that the elements with indices $(a, b)$ and $(b, a)$ are equal to $1$.

Denote by $f_{X_{i}}\left(.\right)$ the density function of $X_{i}$. Similarly, denote by $f_{X_{i},X_{j}}\left(.\right)$ the joint density function of $X_{i}$ and $X_{j}$. The notation $x^{(l)}_{-(i, j)}$ corresponds to the vector $x^{(l)}$ without the $i$-th and $j$-th elements. The notations $x^{(u)}_{-(i, j)}$ and $X_{-(i, j)}$ are defined in the same way.

The cumulative distribution function of the multivariate normal distribution is as follows:
\begin{equation}
F\left(x;\mu,\Sigma\right) = \mathbb{P}\left(X\leq x\right) = \mathbb{P}\left(X_{1}\leq x_{1},...,X_{m}\leq x_{m}\right).
\end{equation}

\section{Examples}\label{sec:examples}

\subsection{Calculation and differentiation of densities}

The function \code{dmnorm} calculates MVN densities $f(x;\mu,\Sigma)$. Its interface is as follows: 

\begin{Sinput}
dmnorm(x, mean, sigma, 
       given_ind     = numeric(),
       log           = FALSE,
       grad_x        = FALSE, 
       grad_sigma    = FALSE,
       is_validation = TRUE, 
       control       = NULL, 
       n_cores       = 1L)
\end{Sinput}

The main arguments of this function are a numeric vector (or matrix) \code{x}, a numeric vector \code{mean}, and a symmetric positive-definite numeric matrix \code{sigma}, representing $x$, $\mu$, and $\Sigma$, respectively. The function returns a list which contains information on the density value (named \code{den}). If \code{grad_x = TRUE}, then the output list also contains the gradient with respect to $x$ named \code{grad_x}. Similarly, if \code{grad_sigma = TRUE}, then the output list additionally contains the gradient with respect to $\Sigma$ named \code{grad_sigma}.

To demonstrate an application of this function, we start with a simple example and gradually move to more advanced cases. Consider a $5$-dimensional MVN vector $X$ with the following parameters:

\begin{align}
\mu &= \left(-2, -1, 0, 1, 2\right),\\
\Sigma &=
\begin{bmatrix}
1 & 0.15 & 0.4 & 0.75 & 1.2\\
0.15 & 2.25 & -0.3 & -0.75 & -1.35\\
0.4 & -0.3 & 4 & 1.5 & 1.2\\
0.75 & -0.75 & 1.5 & 6.25 & -0.375\\
1.2 & -1.35 & 1.2 & -0.375 & 9
\end{bmatrix}.
\end{align}

To calculate $f(x;\mu,\Sigma)$ at the point $x = (-1, 0, 1, 2, 3)$, type:

\begin{Sinput}
library("mnorm")
mean  <- c(-2, -1, 0, 1, 2)
sigma <- matrix(c(1,    0.15,  0.4,  0.75,   1.2,
                  0.15, 2.25,  -0.3, -0.75,  -1.35,
                  0.4,  -0.3,  4,    1.5,    1.2,
                  0.75, -0.75, 1.5,  6.25,   -0.375,
                  1.2,  -1.35, 1.2,  -0.375, 9), 
                ncol = 5)
x     <- c(-1, 0, 1, 2, 3)
d1    <- dmnorm(x = x, mean = mean, sigma = sigma)
\end{Sinput}

Note that the function returns a list whose element named \code{den} contains the density value:
\begin{Sinput}
print(d1$den)
\end{Sinput}
\begin{Soutput}
             [,1]
[1,] 0.0003029506
\end{Soutput}

The function is vectorized with respect to \code{x}. If \code{x} is a matrix, then the density values will be calculated for each row of \code{x}. For example, to additionally calculate the density at the point $y = (2.5, 2, 1.5, 1, 0.5)$, add the following code:

\begin{Sinput}
y     <- c(2.5, 2, 1.5, 1, 0.5)
x_mat <- rbind(x, y)
d2    <- dmnorm(x = x_mat, mean = mean, sigma = sigma)
print(as.vector(d2$den))
\end{Sinput}
\begin{Soutput}
             [,1]
[1] 3.029506e-04 1.254572e-10
\end{Soutput}

Therefore, \code{d2$den[1]} and \code{d2$den[2]} are the calculated values of $f(x;\mu,\Sigma)$ and $f(y;\mu,\Sigma)$, respectively.

To calculate the conditional densities, indices of conditioned elements $I_{g}$ should be provided through the \code{given_ind} argument. Then, elements of \code{x} with corresponding indices will be treated as given values $x^{(g)}$ and other elements will represent $x^{(d)}$. For example, suppose that $X_{1}$ and $X_{3}$ are conditioned at values $-1$ and $2$, respectively, so $x^{(g)}=(-1, 2)$. Then, to calculate the density function of:
\begin{equation}
\left(X_{2}, X_{4}, X_{5}|X_{1} = -1, X_{3} = 2\right),
\end{equation}
at the point $x^{(d)} = (0, 0.5, 1)$, namely:
\begin{equation}
f\left(x^{(d)};\mu,\Sigma | x^{(g)},I_{g}\right)=f\left((0, 0.5, 1);\mu,\Sigma | (-1, 2), (1, 3)\right),
\end{equation}
add the following code:
\begin{Sinput}
z             <- vector(mode = "numeric", length = 5)
given_ind     <- c(1, 3)
x_g           <- c(-1, 2)
x_d           <- c(0, 0.5, 1)
z[given_ind]  <- x_g
z[-given_ind] <- x_d
d3            <- dmnorm(x         = z,     
                        mean      = mean, 
                        sigma     = sigma, 
                        given_ind = given_ind)
print(d3$den)
\end{Sinput}
\begin{Soutput}
            [,1]
[1,] 0.003165719
\end{Soutput}

Note that vectorization with respect to \code{x} is still applicable for the conditional densities. At the same time, the indices of the conditioned elements \code{given_ind} remain identical for all the observations. For example, to additionally calculate the density of:

\begin{equation}
\left(X_{2}, X_{4}, X_{5}|X_{1} = 2.5, X_{3} = 1.5\right),
\end{equation}

at the point $y^{(d)} = (2, 1, 0.5)$, add the code:

\begin{Sinput}
z_mat <- rbind(z, y)
d4    <- dmnorm(x     = z_mat, mean      = mean, 
                sigma = sigma, given_ind = given_ind)
print(as.vector(d4$den))
\end{Sinput}
\begin{Soutput}
[1] 3.165719e-03 3.900875e-05
\end{Soutput}

All aforementioned densities may be differentiated with respect to various parameters. For example, to calculate the gradient of $f(x;\mu,\Sigma)$ with respect to $x$, set \code{grad_x = TRUE} and get the element named \code{grad_x} from the output list:

\begin{Sinput}
d5 <- dmnorm(x     = x,     mean   = mean, 
             sigma = sigma, grad_x = TRUE)
print(signif(d5$grad_x, digits = 3))
\end{Sinput}
\begin{Soutput}
          [,1]      [,2]      [,3]      [,4]      [,5]
[1,] -0.000209 -0.000152 -4.73e-05 -3.16e-05 -2.35e-05
\end{Soutput}

Therefore, \code{d5$grad_x} represents $\nabla f(x;\mu,\Sigma)$. For example, \code{d5$grad_x[3]} is $\frac{\partial f(x;\mu,\Sigma)}{\partial x_{3}}$.

To simultaneously calculate gradients at the points $x$ and $y$ (Jacobian), one may use vectorization:

\begin{Sinput}
d6 <- dmnorm(x     = x_mat, mean   = mean, 
             sigma = sigma, grad_x = TRUE)
print(signif(d6$grad_x, digits = 3))
\end{Sinput}
\begin{Soutput}
          [,1]      [,2]      [,3]      [,4]      [,5]
[1,] -2.09e-04 -1.52e-04 -4.73e-05 -3.16e-05 -2.35e-05
[2,] -7.99e-10  1.82e-12 -5.28e-11  1.17e-10  1.40e-10
\end{Soutput}

Note that \code{grad_x[i, ]} is the gradient at the point \code{x[i, ]}. Particularly, since the first row of \code{x_mat} is $x$ and the second row is $y$, \code{d6$grad_x[1, 3]} is $\frac{\partial f(x;\mu,\Sigma)}{\partial x_{3}}$ and \code{d6$grad_x[2, 5]} is $\frac{\partial f(y;\mu,\Sigma)}{\partial y_{5}}$.

To differentiate the density function with respect to the elements of the covariance matrix, set \code{grad_sigma = TRUE} and get the element named \code{grad_sigma} from the output list:

\begin{Sinput}
d7 <- dmnorm(x    = x_mat, mean       = mean, 
            sigma = sigma, grad_sigma = TRUE)
print(c(d7$grad_sigma[2, 1, 1], d7$grad_sigma[3, 5, 2]))
\end{Sinput}
\begin{Soutput}
[1]  2.101741e-04 -5.358820e-11
\end{Soutput}

Note that the output value \code{grad_sigma} is a $3$-dimensional array such that \code{grad_sigma[, , i]} is the gradient with respect to $\Sigma$ at the point corresponding to \code{x[i, ]}. For example, \code{d7$grad_sigma[2, 1, 1]} is $\frac{\partial f(x;\mu,\Sigma)}{\partial \Sigma_{2,1}}$ and \code{d7$grad_sigma[3, 5, 2]} is $\frac{\partial f(y;\mu,\Sigma)}{\partial \Sigma_{3,5}}$.

Finally, let's differentiate both aforementioned conditional densities:

\begin{Sinput}
d8 <- dmnorm(x          = z_mat, 
             mean       = mean, 
             sigma      = sigma, 
             grad_x     = TRUE,  
             grad_sigma = TRUE, 
             given_ind  = given_ind)
print(signif(d8$grad_x, digits = 3))
\end{Sinput}
\begin{Soutput}
          [,1]      [,2]      [,3]     [,4]     [,5]
[1,] -2.37e-03 -6.93e-05 -6.98e-04 1.41e-03 1.34e-03
[2,] -7.17e-05  5.65e-07 -1.95e-05 3.64e-05 4.34e-05
\end{Soutput}

Note that the output value \code{grad_x} contains derivatives with respect to $x^{(d)}$ and $x^{(g)}$. To distinguish between these two types of derivatives, it is convenient to think of \code{grad_x} as a gradient with respect to the input argument \code{x}. For example, \code{d8$grad_x[1, 3]} is $\frac{\partial f\left(x^{(d)};\mu,\Sigma | x^{(g)},I_{g}\right)}{\partial x^{(g)}_{2}}$ since \code{z_mat[1, 3]} represents the value of $x^{(g)}_{2}$. Similarly, \code{d8$grad_x[2, 5]} is $\frac{\partial f\left(y^{(d)};\mu,\Sigma | y^{(g)},I_{g}\right)}{\partial y^{(d)}_{3}}$ because \code{z_mat[2, 5]} is $y^{(d)}_{3}$. Interpretation of derivatives with respect to $\Sigma$ is identical to the aforementioned case with unconditional densities.

To calculate and differentiate the log-densities, set \code{log = TRUE}. The number of cores to be used for a parallel computation may be set via the \code{n_cores} argument. If \code{validation = FALSE}, then validation of the input arguments will be disabled, which may result in a slight boost in terms of speed. These arguments work similarly for \code{pmnorm}.

\subsection{Calculation and differentiation of probabilities}

The function \code{pmnorm} calculates MVN probabilities $\text{P}\left(x^{(l)}\leq X\leq x^{(u)}\right)$. Its interface is as follows:

\begin{Sinput}
pmnorm(lower, upper,
       given_x            = numeric(),
       mean               = numeric(), 
       sigma              = matrix(),
       given_ind          = numeric(), 
       n_sim              = 1000L,
       method             = "default", 
       ordering           = "mean", 
       log                = FALSE,
       grad_lower         = FALSE, 
       grad_upper         = FALSE, 
       grad_sigma         = FALSE, 
       grad_given         = FALSE, 
       is_validation      = TRUE, 
       control            = NULL, 
       n_cores            = 1L, 
       marginal           = NULL, 
       grad_marginal      = FALSE, 
       grad_marginal_prob = FALSE)
\end{Sinput}

To calculate the bivariate probabilities (by default), the function applies the method of Drezner and Wesolowsky, described in \citep{Genz2004}. In contrast to the classical implementation of this method, \code{pmnorm} applies Gauss-Legendre quadrature with $30$ sample points to approximate the integral (1) of \citep{Genz2004}. The classical implementation of this method uses up to 20 points but requires some additional transformations of (1). During preliminary testing, it has been found that the approach with $30$ points provides similar accuracy, being slightly faster because of better vectorization and parallel computing capabilities.

To calculate the trivariate ($3$-dimensional) probabilities (by default), the function uses the Drezner method, following formula (14) of \citep{Genz2004}. Similarly to the bivariate case, 30 points are used in the Gauss-Legendre quadrature.

For the $m$-variate probabilities, where $m > 3$, the function applies the Geweke-Hajivassiliou-Keane (GHK) algorithm described in section 4.2 of \citep{Genz2009}. The implementation of GHK is based on the deterministic Richtmyer sequence with \code{n_sim} draws and uses variable reordering suggested in Section 4.1.3 of \citep{Genz2009}. The ordering algorithm may be determined via the \code{ordering} argument. Available options are \code{NO}, \code{mean} (default), and \code{variance}. To switch from the Richtmyer sequence to another alternative, the \code{random_sequence} parameter of the \code{control} input argument may be used. A user may supply an $m$-column matrix of uniform (between 0 and 1) numbers generated via some algorithm. For example, the scrambled Halton sequence may be generated via the \code{halton} function of the \code{mnorm} package. The bivariate and trivariate probabilities may also be calculated via the GHK algorithm (not recommended in terms of speed and accuracy) by setting \code{method = "GHK"}.

If \code{method = "Gassmann"}, then the function applies the method of \citep{Gassmann2003} for the calculation of the $m>3$-dimensional normal probabilities. It uses the matrix representation (5) of \citep{Gassmann2003} and $30$ points of the Gauss-Legendre quadrature.

The main arguments are \code{lower}, \code{upper}, \code{mean}, and \code{sigma} representing $x^{(l)}$, $x^{(u)}$, $\mu$, and $\Sigma$, respectively. Note that \code{pmnorm} is optimized to perform much faster (especially in the $2$- and $3$-dimensional cases) when some integration limits are infinite. The function returns a list which contains information on the probability value (named \code{prob}). The list may also contain the derivatives named after the gradient-related arguments \code{grad_lower}, \code{grad_upper}, \code{grad_sigma}, \code{grad_given}, and \code{grad_marginal}. For example, if \code{grad_upper = TRUE}, then the output list will contain the element named \code{grad_upper} which stores the information on the derivatives with respect to the upper integration limits. Let's clarify the application of the \code{pmnorm} function and the structure of its output by investigating a series of pedagogical examples.

Consider a $5$-dimensional MVN vector from the previous section. Let's calculate the following probability:
\begin{gather}
\mathbb{P}(x^{(l)} < X < x^{(u)}) =  \mathbb{P}(-3 < X_{1} < 1, -2.5 < X_{2} < 1.5,\\
    -2 < X_{3} < 2, -1.5 < X_{4} < 2.5, -1 < X_{5} < 3),\\
x^{(l)} = \left(-3, -2.5, -2, -1.5, -1\right), \quad
x^{(u)} = \left(1, 1.5, 2, 2.5, 3\right).
\end{gather}

To calculate this probability, run the following code:
\begin{Sinput}
library("mnorm")
mean  <- c(-2, -1, 0, 1, 2)
sigma <- matrix(c(1,    0.15,  0.4,  0.75,   1.2,
                  0.15, 2.25,  -0.3, -0.75,  -1.35,
                  0.4,  -0.3,  4,    1.5,    1.2,
                  0.75, -0.75, 1.5,  6.25,   -0.375,
                  1.2,  -1.35, 1.2,  -0.375, 9), 
                ncol = 5)
lower <- c(-3, -2.5, -2, -1.5, -1)
upper <- c(1,   1.5,  2,  2.5,  3)
p1    <- pmnorm(lower = lower, upper = upper, 
                mean  = mean,  sigma = sigma)
print(p1$prob)
\end{Sinput}
\begin{Soutput}
          [,1]
[1,] 0.1326377
\end{Soutput}

The function is vectorized with respect to the \code{lower} and \code{upper} arguments. If these arguments are matrices, then the limits should be provided in the rows. Let's additionally calculate the probability with the following limits:
\begin{gather}
y^{(l)} = \left(-\infty, 0, 1, 2, -\infty\right), \quad
y^{(u)} = \left(-1, \infty, 2, \infty, 1\right).
\end{gather}

To simultaneously calculate both aforementioned probabilities, add the code:
\begin{Sinput}
lower_mat <- rbind(lower, c(-Inf,  0,   1,   2, -Inf))
upper_mat <- rbind(upper, c(   -1, Inf, 2, Inf, 1))
p2        <- pmnorm(lower = lower_mat, upper = upper_mat, 
                    mean  = mean,      sigma = sigma)
print(as.vector(p2$prob))
\end{Sinput}
\begin{Soutput}
[1] 0.132637686 0.004791223
\end{Soutput}

Therefore, \code{p2$prob[1]} and \code{p2$prob[2]} are the calculated values of the MVN probabilities under the limits $x^{(l)}$, $x^{(u)}$ and $y^{(l)}$, $y^{(u)}$, respectively.

Let's calculate the conditional probabilities:
\begin{gather}
P_{1} = \mathbb{P}\left(x^{(l)} < X_{d} < x^{(u)} | X_{g} = x^{(g)}\right) =\\
=\mathbb{P}(-2.5 < X_{2} < 1.5, -1.5 < X_{4} < 2.5, -1 < X_{5} < 3 | X_{1} = 1, X_{3} = -2),
\end{gather}
and:
\begin{gather}
P_{2} = \mathbb{P}\left(y^{(l)} < X_{d} < y^{(u)} | X_{g} = y^{(g)}\right) =\\
=\mathbb{P}(0 < X_{2} < \infty, 2 < X_{4} < \infty, -\infty < X_{5} < 1 | X_{1} = -1, X_{3} = 2),
\end{gather}
where:
\begin{gather}
x^{(l)} = \left(-2.5, -1.5, -1\right), x^{(u)} = \left(1.5, 2.5, 3\right),\\
y^{(l)} = \left(0, 2, -\infty\right), y^{(u)} = \left(\infty, \infty, 1\right),\\
x^{(g)} = \left(1, -2\right),\quad y^{(g)} = \left(-1, 2\right), I_{g}=\left(1, 3\right).
\end{gather}

The conditional values $x^{(g)}$ and $y^{(g)}$ should be provided as the rows of the vectorized argument \code{given_x}. The indices of the conditioned components $I_{g}$ should be supplied to the \code{given_ind} argument. Similarly to the previous example, $x^{(l)}, x^{(u)}$ and $y^{(l)}, y^{(u)}$ should be the rows of the \code{lower} and \code{upper} arguments, respectively. Therefore, to calculate these conditional probabilities, add the following code:
\begin{Sinput}
lower_d <- rbind(c(-2.5, -1.5, -1), c(0,   2,   -Inf))
upper_d <- rbind(c( 1.5,  2.5,  3), c(Inf, Inf,  1))
p3      <- pmnorm(lower     = lower_d, 
                  upper     = upper_d, 
                  mean      = mean, 
                  sigma     = sigma,
                  given_ind = c(1, 3), 
                  given_x   = rbind(c(1, -2), c(-1, 2)))
print(as.vector(p3$prob))
\end{Sinput}
\begin{Soutput}
[1] 0.06055756 0.04535331
\end{Soutput}

The arguments \code{grad_lower}, \code{grad_upper}, \code{grad_sigma}, and \code{grad_given} allow one to differentiate the aforementioned conditional probabilities with respect to $x^{(l)}$ and $y^{(l)}$, $x^{(u)}$ and $y^{(u)}$, $\Sigma$, and $x^{(g)}$ and $y^{(g)}$, respectively:
\begin{Sinput}
p4 <- pmnorm(lower      = lower_d, 
             upper      = upper_d, 
             mean       = mean,
             sigma      = sigma,
             given_ind  = c(1, 3), 
             given_x    = rbind(c(1, -2), c(-1, 2)),
             grad_lower = TRUE, 
             grad_upper = TRUE,
             grad_sigma = TRUE, 
             grad_given = TRUE)
print(c(p4$grad_lower[1, 2], p4$grad_upper[2, 3], 
        p4$grad_given[2, 1], p4$grad_sigma[4, 5, 1]))
\end{Sinput}
\begin{Soutput}
[1] -0.003092860  0.016916924 -0.003887902  0.012928930
\end{Soutput}

Note that \code{grad_lower[i, ]}, \code{grad_upper[i, ]}, and \code{grad_given[i, ]} are the gradients of the $i$-th probability with respect to the lower integration limits, the upper integration limits, and the conditioned values, respectively. For example, \code{grad_lower[1, 2]} is $\frac{\partial P_{1}}{\partial x^{(l)}_{2}}$, \code{grad_upper[2, 3]} is $\frac{\partial P_{2}}{\partial y^{(u)}_{3}}$, and \code{grad_given[2, 1]} is $\frac{\partial P_{2}}{\partial y^{(g)}_{1}}$. The value of \code{grad_sigma[j, k, i]} represents the derivative of the $i$-th probability with respect to $\Sigma_{j, k}$. Particularly, \code{grad_sigma[4, 5, 1]} is $\frac{\partial P_{1}}{\partial \Sigma_{4, 5}}$.

Finally, let's substitute the marginal distributions using the \code{marginal} argument. It should be a list such that \code{names(marginal)[i]} is the name of the $i$-th distribution and \code{marginal[[i]]} represents its parameters. Possible names (distributions) are \code{"PGN"} (or \code{"hpa"}), \code{"student"} (or \code{"t"}), \code{"logistic"}, and \code{"normal"}. Normal and logistic distributions should take \code{NULL} or an empty vector \code{numeric()} as the parameter. For the Student distribution, it should be a single value representing the degrees of freedom. To incorporate the PGN distribution proposed by \citep{PGN}, we use the \code{phpa0} function from the \textbf{hpa} package, allowing one to calculate the probabilities of the standardized PGN distribution as described in \citep{Trifonov}. Therefore, the values from the corresponding elements of \code{marginal} are transferred to the argument \code{pc} of the \code{phpa0} function.

Suppose that $X_{1}$ and $X_{2}$ have the PGN distribution (with some parameters), $X_{3}$ has the Student distribution with $5$ degrees of freedom, $X_{4}$ has the standard logistic distribution, and $X_{5}$ has the normal distribution. The distribution of each $X_{i}$ is automatically transformed to have mean $\mu_{i}$ and variance $\Sigma_{i,i}$. To calculate the probabilities given these marginal distributions (connected by the Gaussian copula), run the following code:

\begin{Sinput}
p5 <- pmnorm(lower          = lower_d, 
             upper          = upper_d,
             mean           = mean, 
             sigma          = sigma,
             given_ind      = c(1, 3), 
             given_x        = rbind(c(1, -2), c(-1, 2)),
             grad_lower     = TRUE, 
             grad_upper     = TRUE,
             grad_sigma     = TRUE, 
             grad_given     = TRUE,
             grad_marginal  = TRUE,
             marginal       = list(
               PGN      = c(0.5, -0.2, 0.1), 
               hpa      = c(0.2, 0.05),
               student  = 5, 
               logistic = numeric(), 
               normal   = NULL))
print(p5$grad_marginal[[1]])
\end{Sinput}
\begin{Soutput}
              [,1]          [,2]         [,3]
[1,] -0.0389503917  0.0106233070 -0.236806340
[2,] -0.0002633928 -0.0001882154  0.007446111
\end{Soutput}

By setting \code{grad_marginal = TRUE}, we additionally calculate the derivatives of the probabilities with respect to the parameters of the marginal distributions. Since normal and logistic distributions have no parameters, \code{p5$grad_marginal[[4]]} and \code{p5$grad_marginal[[5]]} are empty. The value of \code{p5$grad_marginal[[3]][i]} represents the derivative of the $i$-th probability with respect to the degrees of freedom of the Student distribution. Finally, \code{p5$grad_marginal[[1]][i, ]} and \code{p5$grad_marginal[[2]][i, ]} contain the gradient of the $i$-th probability with respect to the parameters of the PGN distributions associated with $X_{1}$ and $X_{2}$, respectively.

\section{Efficiency analysis}\label{sec:efficiency}

\subsection{Derivatives}

To motivate the application of the analytical derivatives, we compare their computation speed (in seconds) to that of numeric derivatives. The numeric gradient requires at least one more call to the function than the number of its elements. For example, to get the numeric gradient of $f(x)$ with respect to $x$ in the $m=5$-dimensional case, the numeric gradient requires at least $m+1=6$ calls to $f(x)=f(x_{1},...,x_{5})$. We will use this fact to calculate a lower bound of the time needed to compute the numeric gradient. Note that accurate numeric differentiation techniques may require $2$ or more calls to the function per parameter. Also, this lower bound ignores the time needed to combine the calls to the function into a gradient. Nevertheless, the following results suggest that in most cases the calculation time of the analytical derivatives of the MVN densities and probabilities notably outperforms this lower bound.

Consider the same MVN distribution setup as in the previous sections. First, let's compare the calculation time of the densities, the conditional densities, and their derivatives on $10^6$ simulations.

\begin{Sinput}
library("mnorm")
set.seed(123)
mean  <- c(-2, -1, 0, 1, 2)
sigma <- matrix(c(1,    0.15,  0.4,  0.75,   1.2,
                  0.15, 2.25,  -0.3, -0.75,  -1.35,
                  0.4,  -0.3,  4,    1.5,    1.2,
                  0.75, -0.75, 1.5,  6.25,   -0.375,
                  1.2,  -1.35, 1.2,  -0.375, 9), 
                ncol = 5)
m         <- length(mean)
n_sigma   <- (m ^ 2 + m) / 2
x         <- rmnorm(10 ^ 6, mean  = mean, 
                            sigma = sigma)
given_ind <- c(1, 3)
den       <- system.time(dmnorm(x     = x, 
                                mean  = mean, 
                                sigma = sigma))[[3]]
den_cond  <- system.time(dmnorm(x         = x, 
                                mean      = mean, 
                                sigma     = sigma, 
                                given_ind = given_ind))[[3]]
grad_x          <- system.time(
  dmnorm(x      = x,         
         mean   = mean, 
         sigma  = sigma, 
         grad_x = TRUE))[[3]]
grad_sigma      <- system.time(
  dmnorm(x          = x,    
         mean       = mean, sigma = sigma, 
         grad_x     = TRUE, 
         grad_sigma = TRUE))[[3]]
grad_x_cond     <- system.time(
  dmnorm(x        = x, 
        mean      = mean, 
        given_ind = given_ind,
        sigma     = sigma, 
        grad_x    = TRUE))[[3]]
grad_sigma_cond <- system.time(
  dmnorm(x          = x, 
         mean       = mean, 
         sigma      = sigma, 
         grad_x     = TRUE, 
         grad_sigma = TRUE,
         given_ind  = given_ind))[[3]]
\end{Sinput}

The computation time for $f(x)$ and $f(x^{(d)}; x^{(g)})$, denoted by \code{den_uncond} and \code{den_cond}, respectively (similarly for the following derivatives of these densities), is as follows:

\begin{Sinput}
print(c(den_uncond = den, den_cond = den_cond))
\end{Sinput}
\begin{Soutput}
den_uncond   den_cond 
      0.09       0.12
\end{Soutput}

The computation time of the gradients with respect to $x$ (unconditional) and $x^{(d)}$, $x^{(g)}$ (conditional), where prefix \code{num} stands for the computed value of the lower bound of time needed to calculate the numeric derivatives (similarly for further examples), is as follows:

\begin{Sinput}
print(c(grad_x_uncond     = grad_x, 
        grad_x_cond       = grad_x_cond, 
        grad_x_uncond_num = den * (m + 1), 
        grad_x_cond_num   = den_cond * (m + 1)))
\end{Sinput}
\begin{Soutput}
    grad_x_uncond       grad_x_cond 
             0.09              0.12 
grad_x_uncond_num   grad_x_cond_num 
             0.54              0.72
\end{Soutput}

The computation time of the gradients with respect to all the parameters, i.e., $x$, $x^{(d)}$, $x^{(g)}$, and $\Sigma$, is as follows:

\begin{Sinput}
print(c(grad_sigma_uncond     = grad_sigma, 
        grad_sigma_cond       = grad_sigma_cond, 
        grad_sigma_uncond_num = den * (n_sigma + m + 1),
        grad_sigma_cond_num   = den_cond * (n_sigma + m + 1)))
\end{Sinput}
\begin{Soutput}
    grad_sigma_uncond       grad_sigma_cond 
                 0.70                  1.60 
grad_sigma_uncond_num   grad_sigma_cond_num 
                 1.89                  2.52 
\end{Soutput}

The results suggest that computation time for the unconditional and conditional densities is just slightly greater than that of their derivatives with respect to $x$, $x^{(d)}$, and $x^{(g)}$. As a result, analytical derivatives with respect to these parameters are much faster than numeric derivatives. In contrast, the time needed to calculate the analytical derivatives of the (conditional) densities with respect to all the parameters (including elements of $\Sigma$) is just mildly smaller than the minimum time needed to compute the numeric derivatives of the (conditional) densities. It may be associated with the fact that the analytical derivatives with respect to $\Sigma$ are rather complicated, especially in the conditional case.

Let's use $10^3$ simulations to calculate the efficiency of the analytical gradients for the unconditional probabilities.

\begin{Sinput}
set.seed(123)
lower <- matrix(-Inf, nrow = 10 ^ 3, ncol = 5)
upper <- rmnorm(10 ^ 3, mean = mean, sigma = sigma)
prob       <- system.time(
  pmnorm(lower = lower, 
         upper = upper, 
         mean  = mean,  
         sigma = sigma))[[3]]
grad_upper <- system.time(
  pmnorm(lower      = lower,
         upper      = upper, 
         mean       = mean,  
         sigma      = sigma, 
         grad_upper = TRUE))[[3]]
grad_sigma <- system.time(
  pmnorm(lower      = lower, 
         upper      = upper, 
         mean       = mean, 
         sigma      = sigma, 
         grad_upper = TRUE, 
         grad_sigma = TRUE))[[3]]
print(c(prob           = prob, 
        grad_upper     = grad_upper, 
        grad_sigma     = grad_sigma,
        grad_upper_num = prob * (m + 1), 
        grad_sigma_num = prob * (n_sigma + m + 1)))
\end{Sinput}
\begin{Soutput}
          prob     grad_upper     grad_sigma 
          0.14           0.61           0.64 
grad_upper_num grad_sigma_num 
          0.84           2.94
\end{Soutput}

The results indicate that the analytical gradients are much faster than the numeric gradients for $\Sigma$ and just slightly faster for $x^{(u)}$. It may be explained by the fact that the derivatives of the probabilities with respect to $x^{(u)}$ require the calculation of the $(m-1)$-dimensional integrals, while the derivatives with respect to $\Sigma$ require the calculation of the $(m-2)$-dimensional integrals. Notice that in the $5$-dimensional case, the $4$-dimensional integrals are still calculated via the GHK algorithm, which is much slower than the Drezner method used for the $3$-dimensional probabilities. This explains why the extra time needed to calculate the derivatives with respect to $\Sigma$ is so small.

Let's reproduce the experiment for the conditional probabilities using $10^5$ simulations.

\begin{Sinput}
set.seed(123)
given_ind <- c(1, 3)
m_g       <- length(given_ind)
lower     <- matrix(-Inf, nrow = 10 ^ 5, ncol = m - m_g)
x         <- rmnorm(10 ^ 5, mean = mean, sigma = sigma)
upper     <- x[, -given_ind]
given_x   <- x[, given_ind]
prob <- system.time(
  pmnorm(lower     = lower,
         upper     = upper, 
         mean      = mean, 
         sigma     = sigma,
         given_ind = given_ind, 
         given_x   = given_x))[[3]]
grad_upper <- system.time(
  pmnorm(lower      = lower, 
         upper      = upper,
         given_ind  = given_ind,
         given_x    = given_x,
         mean       = mean, 
         sigma      = sigma, 
         grad_upper = TRUE))[[3]]
grad_given <- system.time(
  pmnorm(lower      = lower , 
         upper      = upper,
         given_ind  = given_ind, 
         given_x    = given_x,
         mean       = mean, 
         sigma      = sigma, 
         grad_upper = TRUE, 
         grad_given = TRUE))[[3]]
grad_sigma <- system.time(
  pmnorm(lower      = lower, 
         upper      = upper, 
         mean       = mean, 
         sigma      = sigma,
         given_ind  = given_ind, 
         given_x    = given_x,
         grad_upper = TRUE, 
         grad_given = TRUE,
         grad_sigma = TRUE))[[3]]
print(c(prob           = prob, 
        grad_upper     = grad_upper, 
        grad_given     = grad_given, 
        grad_sigma     = grad_sigma, 
        grad_upper_num = prob * (m - m_g + 1),
        grad_given_num = prob * (m + 1), 
        grad_sigma_num = prob * (n_sigma + m + 1)))
\end{Sinput}
\begin{Soutput}
          prob     grad_upper     grad_given 
          0.33           0.52           0.52 
    grad_sigma grad_upper_num grad_given_num 
          0.69           1.32           1.98 
grad_sigma_num 
          6.93
\end{Soutput}

For the conditional probabilities, the difference in the calculation speed between analytical and numeric derivatives is more substantial. This is because computation of the $5$-dimensional conditional MVN probabilities requires the $m_{d}=3$-dimensional integration (since $2$ out of $5$ components are conditioned). Therefore, the derivatives are calculated extremely fast via the $1$-dimensional and $2$-dimensional integration methods. Finally, let's consider the $10$-dimensional case when all the derivatives will be calculated using the probabilities calculated via the GHK method. Let's use $10^3$ simulations, a zero mean vector and a constant correlation matrix with the correlation coefficients equal to $0.5$.

\begin{Sinput}
set.seed(123)
m           <- 10
n_sigma     <- (m ^ 2 + m) / 2
mean        <- rep(0, m)
sigma       <- matrix(0.5, nrow = m, ncol = m)
diag(sigma) <- 1
lower       <- matrix(-Inf, nrow = 10 ^ 3, ncol = m)
upper       <- rmnorm(10 ^ 3, mean = mean, sigma = sigma)
prob        <- system.time(
  pmnorm(lower = lower, 
         upper = upper, 
         mean  = mean, 
         sigma = sigma))[[3]]
grad_upper  <- system.time(
  pmnorm(lower      = lower,
         upper      = upper, 
         mean       = mean, 
         sigma      = sigma, 
         grad_upper = TRUE))[[3]]
grad_sigma  <- system.time(
  pmnorm(lower      = lower, 
         upper      = upper, 
         mean       = mean, 
         sigma      = sigma, 
         grad_upper = TRUE, 
         grad_sigma = TRUE))[[3]]
print(c(prob           = prob, 
        grad_upper     = grad_upper, 
        grad_sigma     = grad_sigma,
        grad_upper_num = prob * (m + 1), 
        grad_sigma_num = prob * (n_sigma + m + 1)))
\end{Sinput}
\begin{Soutput}
          prob     grad_upper     grad_sigma 
          0.41           3.47          14.55 
grad_upper_num grad_sigma_num 
          4.51          27.06
\end{Soutput}

In line with the aforementioned expectations, the results demonstrate a substantial decline in the relative performance of the analytical derivatives. Nevertheless, their calculation time is still slightly less than the lower bound for the numeric derivatives.

Finally, notice that the derivatives with respect to $x^{(g)}$ and $\Sigma_{i,j}$ (such that $i\in I_{g}$ or $j\in I_{g}$) are just functions of the derivatives with respect to $x^{(u)}$ and $\Sigma_{k, l}$ (such that $k,l\in I_{d}$). Hence, extra time needed to calculate these derivatives is very low. Consequently, the aforementioned decline should be much lower for the conditional probabilities. Indeed, suppose that $I_{g}=\{1,2,3,4\}$. Then both the absolute and the relative (to the numeric gradient) time to calculate the analytical derivatives with respect to $x^{(u)}$, $x^{(g)}$ and $\Sigma$ decrease substantially.

\begin{Sinput}
set.seed(123)
given_ind <- 1:4
m_g       <- length(given_ind)
lower     <- matrix(-Inf, nrow = 10 ^ 3, ncol = m - m_g)
x         <- rmnorm(10 ^ 3, mean = mean, sigma = sigma)
upper     <- x[, -given_ind]
given_x   <- x[, given_ind]
prob <- system.time(
  pmnorm(lower     = lower, 
         upper     = upper, 
         mean      = mean, 
         sigma     = sigma,
         given_ind = given_ind, 
         given_x   = given_x))[[3]]
grad_upper <- system.time(
  pmnorm(lower      = lower, 
         upper      = upper, 
         mean       = mean, 
         sigma      = sigma,
         given_ind  = given_ind, 
         given_x    = given_x,
         grad_upper = TRUE))[[3]]
grad_given <- system.time(
  pmnorm(lower      = lower, 
         upper      = upper, 
         mean       = mean, 
         sigma      = sigma,
         given_ind  = given_ind, 
         given_x    = given_x,
         grad_upper = TRUE, 
         grad_given = TRUE))[[3]]
grad_sigma <- system.time(
  pmnorm(lower      = lower, 
         upper      = upper, 
         mean       = mean, 
         sigma      = sigma, 
         given_ind  = given_ind, 
         given_x    = given_x,
         grad_upper = TRUE, 
         grad_given = TRUE,
         grad_sigma = TRUE))[[3]]
print(c(prob           = prob, 
        grad_upper     = grad_upper, 
        grad_given     = grad_given, 
        grad_sigma     = grad_sigma,
        grad_upper_num = prob * (m - m_g + 1), 
        grad_given_num = prob * (m + 1), 
        grad_sigma_num = prob * (n_sigma + m + 1)))
\end{Sinput}
\begin{Soutput}
          prob     grad_upper     grad_given 
          0.16           0.90           0.89 
    grad_sigma grad_upper_num grad_given_num 
          2.23           1.12           1.76 
grad_sigma_num 
         10.56
\end{Soutput}

\subsection{Densities}

We compare the efficiency of MVN density computation by \textbf{mnorm} with that of the \textbf{mvnfast} and \textbf{mvtnorm} packages, which calculate these densities via the functions \code{dmvn} and \code{dmvnorm}, respectively. To do so, we simulate $10$ million observations from the $5$-dimensional MVN distribution. Then we compute the MVN density value at each of the simulated points and calculate computation time (in seconds) for each function. Since both \code{dmnorm} and \code{dmvn} support parallel computing, we also compare these functions using $8$ cores.
 
\begin{Sinput}
library("mnorm")
library("mvnfast")
library("mvtnorm")
set.seed(123)
n     <- 10 ^ 7
mean  <- c(-2, -1, 0, 1, 2)
sigma <- matrix(c(1,    0.15,  0.4,  0.75,   1.2,
                  0.15, 2.25,  -0.3, -0.75,  -1.35,
                  0.4,  -0.3,  4,    1.5,    1.2,
                  0.75, -0.75, 1.5,  6.25,   -0.375,
                  1.2,  -1.35, 1.2,  -0.375, 9), ncol = 5)
x     <- rmnorm(n, mean = mean, sigma = sigma)
cbind(mnorm            = system.time(
        dmnorm(x     = x, 
               mean  = mean, 
               sigma = sigma))[[3]],
      mvnfast          = system.time(
        dmvn(X     = x, 
             mu    = mean, 
             sigma = sigma))[[3]],
      mnorm_parallel   = system.time(
        dmnorm(x       = x, 
               mean    = mean, 
               sigma   = sigma, 
               n_cores = 8))[[3]],
      mvnfast_parallel = system.time(
        dmvn(X      = x, 
             mu     = mean, 
             sigma  = sigma, 
             ncores = 8))[[3]],
      mvtnorm          = system.time(
        dmvnorm(x     = x, 
                mean  = mean, 
                sigma = sigma))[[3]])
\end{Sinput}
\begin{Soutput}
     mnorm mvnfast mnorm_parallel
[1,]   0.5    0.61           0.23
     mvnfast_parallel mvtnorm
[1,]              0.3    0.75
\end{Soutput}

The results suggest that both \code{dmnorm} and \code{dmvn} notably outperform \code{dmvnorm}.
Also, under the same number of cores, \code{dmnorm} is faster than \code{dmvn}.

Let's replicate the example for the $100$-dimensional case with a zero mean vector and a constant correlation matrix whose correlation coefficients equal $0.5$. We consider $10^5$ simulations.

\begin{Sinput}
set.seed(123)
m           <- 100
n           <- 10 ^ 5
mean        <- rep(0, m)
sigma       <- matrix(0.5, nrow = m, ncol = m)
diag(sigma) <- 1
x           <- rmnorm(n, mean = mean, sigma = sigma)
cbind(mnorm            = system.time(
        dmnorm(x     = x, 
               mean  = mean, 
               sigma = sigma))[[3]],
      mvnfast          = system.time(
        dmvn(X     = x, 
             mu    = mean, 
             sigma = sigma))[[3]],
      mnorm_parallel   = system.time(
        dmnorm(x       = x, 
               mean    = mean, 
               sigma   = sigma, 
               n_cores = 8))[[3]],
      mvnfast_parallel = system.time(
        dmvn(X      = x, 
             mu     = mean, 
             sigma  = sigma, 
             ncores = 8))[[3]],
      mvtnorm          = system.time(
        dmvnorm(x     = x, 
                mean  = mean, 
                sigma = sigma))[[3]])
\end{Sinput}
\begin{Soutput}
     mnorm mvnfast mnorm_parallel
[1,]   0.4    0.39           0.14
     mvnfast_parallel mvtnorm
[1,]              0.1    0.39
\end{Soutput}

The results suggest that, in the high-dimensional case, all the non-parallel functions provide similar computation speed. However, the parallel implementations of \code{dmnorm} and \code{dmvn} substantially outperform \code{dmvnorm}.

\subsection{Bivariate and trivariate probabilities}

First, we compare the speed of the \code{pmnorm} and \code{pbivnorm} functions using the standard bivariate normal distribution with correlation coefficients equal to $0.2$, $0.5$, $0.8$, and $0.95$.

\begin{Sinput}
library("mnorm")
library("pbivnorm")
set.seed(123)
mean <- c(0, 0)
cors <- c(0.2, 0.5, 0.8, 0.95)
ad   <- rep(0, length(cors))
rd   <- ad
for (i in 1:length(cors))
{
  sigma   <- matrix(c(1, cors[i], cors[i], 1), ncol = 2)
  x       <- rmnorm(10 ^ 6, mean = mean, sigma = sigma)
  neg_inf <- matrix(-Inf, nrow = 10 ^ 6, ncol = 2)
  cat(c("The correlation coefficient equals ", cors[i], "\n"))
  print(
    cbind(mnorm = system.time(
            p1 <- pmnorm(lower = neg_inf, 
                         upper = x, 
                         mean  = mean, 
                         sigma = sigma)$prob)[[3]],
          mnorm_parallel = system.time(
            p2 <- pmnorm(lower   = neg_inf, 
                         upper   = x, 
                         mean    = mean, 
                         sigma   = sigma, 
                         n_cores = 8)$prob)[[3]],
          pbivnorm = system.time(
            p3 <- pbivnorm(x   = x[, 1], 
                           y   = x[, 2], 
                           rho = cors[i]))[[3]])
  )
  cat("\n")
  ad[i] <- max(abs(p1 - p3))
  rd[i] <- max(abs(p1 / p3) - 1)
}
\end{Sinput}
\begin{Soutput}
The correlation coefficient equals  0.2 
     mnorm mnorm_parallel pbivnorm
[1,]  0.57           0.16     0.75

The correlation coefficient equals  0.5 
     mnorm mnorm_parallel pbivnorm
[1,]  0.57           0.14     1.11

The correlation coefficient equals  0.8 
     mnorm mnorm_parallel pbivnorm
[1,]  0.56           0.15      1.8

The correlation coefficient equals  0.95 
     mnorm mnorm_parallel pbivnorm
[1,]  0.56           0.14     1.93
\end{Soutput}

The results suggest that, independent of the correlation coefficient value, \code{pmnorm} is notably faster than \code{pbivnorm}. To check that the accuracy of both functions is similar, we also calculate the maximum absolute and relative differences between the probabilities computed via \code{pmnorm} and \code{pbivnorm}.

\begin{Sinput}
for (i in 1:length(cors))
{
  cat(c("The correlation coefficient equals ", cors[i], "\n"))
  print(c(absolute_difference = ad[i], 
          relative_difference = rd[i]))
  cat("\n")
}
\end{Sinput}
\begin{Soutput}
The correlation coefficient equals  0.2 
absolute_difference relative_difference 
       5.551115e-16        7.172041e-14 

The correlation coefficient equals  0.5 
absolute_difference relative_difference 
       4.440892e-16        8.881784e-16 

The correlation coefficient equals  0.8 
absolute_difference relative_difference 
       4.440892e-16        1.332268e-15 

The correlation coefficient equals  0.95 
absolute_difference relative_difference 
       2.059464e-14        5.886402e-13
\end{Soutput}

The accuracy of both functions is very similar since the maximum absolute and relative differences in the calculated probabilities are extremely small.

To our knowledge, the fastest way to compute the trivariate probabilities in \textbf{R} is by using the \textbf{TVPACK} algorithm from the \code{mvtnorm} package. Therefore, we compare our algorithm to this function on $20000$ simulations from the standard trivariate normal distribution with the correlation coefficients $0.8$, $0.7$, and $0.6$. Since \code{mvtnorm} is not vectorized, we will use a loop to calculate the corresponding probabilities. 

\begin{Sinput}
library("mnorm")
library("mvtnorm")
set.seed(123)
mean    <- c(0, 0, 0)
sigma   <- matrix(c(1,    0.8,  0.7,
                    0.8,  1,    0.6,
                    0.7,  0.6,  1), ncol = 3)
n       <- 20000
x       <- rmnorm(n, mean = mean, sigma = sigma)
neg_inf <- matrix(-Inf, nrow = n, ncol = 3)
cbind(mnorm = system.time(
        pmnorm(lower = neg_inf,
               upper = x, 
               mean  = mean, 
               sigma = sigma))[[3]],
      mnorm_parallel = system.time(
        pmnorm(lower   = neg_inf, 
               upper   = x, 
               mean    = mean, 
               sigma   = sigma, 
               n_cores = 8))[[3]],
      pmvnorm = system.time(
      for (i in 1:n)
      {
        pmvnorm(lower     = neg_inf[1, ], 
                upper     = x[i, ], 
                mean      = mean, 
                sigma     = sigma, 
                algorithm = TVPACK())
      })[[3]])
\end{Sinput}
\begin{Soutput}
     mnorm mnorm_parallel pmvnorm
[1,]  0.06           0.01    4.28
\end{Soutput}

The results suggest that \code{pmnorm} dramatically outperforms \code{pmvnorm}. We have additionally checked that the difference in performance remains very similar even for large values of the \code{abseps} input argument of \code{TVPACK} (omitted for brevity). To ensure that the results are not driven by substantial accuracy losses, we have also checked that the absolute and relative differences in the calculated probabilities are small.

\begin{Sinput}
p1 <- pmnorm(lower = neg_inf, 
             upper = x,
             mean  = mean, 
             sigma = sigma)$prob
p2 <- vector(mode = "numeric", length = n)
for (i in 1:n)
{
  p2[i] <- pmvnorm(lower     = neg_inf[1, ], 
                   upper     = x[i, ], 
                   mean      = mean, 
                   sigma     = sigma, 
                   algorithm = TVPACK())
}
c(absolute_difference = max(abs(p1 - p2)), 
  relative_difference = max(abs(p1 / p2 - 1)))
\end{Sinput}
\begin{Soutput}
absolute_difference relative_difference 
       4.440892e-16        2.553513e-15
\end{Soutput}

\subsection{High-dimensional probabilities}

To test the efficiency of \code{pmnorm} in the high-dimensional case, we have replicated Tables 1-3 of \citep{tlrmvnmvt} (using their code from the supplementary materials) for $16$, $64$, $128$, and $512$ dimensions. Our modified version of this code may be provided upon request. All tables have a constant correlation matrix. For Table 1 and Table 2, the correlation coefficients equal $0.5$. For Table 3 they equal $0.8$. All lower integration limits are $-\infty$. For Table 2 and Table 3, the upper integration limits are $-1$. For Table 1 they equal $0$. Following \citep{tlrmvnmvt}, we use the default settings for all the functions. For \code{pmnorm}, we set \code{n_sim = 20000} since with this value its calculation time is similar to that of other functions. Columns of these tables represent the number of dimensions. Rows correspond to the functions or packages used to calculate the probabilities. The row labeled TruncatedNormal stands for the \code{pmvnorm} function from the \textbf{TruncatedNormal} package. The rows labeled tlrmvnmvt (GenzBretz) and tlrmvnmvt (TLRQMC) represent the function \code{pmvn} from \textbf{tlrmvnmvt} with the \code{algorithm} argument set to \code{GenzBretz} and \code{TLRQMC}, respectively.

% Table 1
\begin{table}[htbp]
\footnotesize
  \centering
  \caption{Correlation 0.5, upper integration limits 0.}
  \label{tab:1}
  \begin{tabular}{lcccc}
    \toprule
    & 16 & 64 & 128 & 512 \\
    \midrule
    mvtnorm                & 0.01\% (0.03s) & 0.07\% (0.20s) & 0.17\% (0.47s) & 0.53\% (3.47s) \\
    TruncatedNormal        & 0.15\% (0.04s) & 0.08\% (0.28s) & 0.13\% (0.85s) & 0.11\% (11.52s) \\
    tlrmvnmvt (GenzBretz)  & 0.05\% (0.02s) & 0.21\% (0.07s) & 0.38\% (0.15s) & 0.74\% (0.87s) \\
    tlrmvnmvt (TLRQMC)     & 0.05\% (0.02s) & 0.19\% (0.09s) & 0.37\% (0.18s) & 0.74\% (0.79s) \\
    pmnorm (mnorm)         & 0.06\% (0.01s) & 0.11\% (0.07s) & 0.23\% (0.20s) & 0.44\% (2.54s) \\
    \bottomrule
  \end{tabular}
\end{table}

% Table 2
\begin{table}[htbp]
\footnotesize
  \centering
  \caption{Correlation 0.5, upper integration limits -1.}
  \label{tab:2}
  \begin{tabular}{lcccc}
    \toprule
    & 16 & 64 & 128 & 512 \\
    \midrule
    mvtnorm                & 0.03\% (0.03s) & 0.32\% (0.18s) & 0.74\% (0.44s) & 2.84\% (3.32s) \\
    TruncatedNormal        & 0.06\% (0.05s) & 0.07\% (0.28s) & 0.09\% (0.85s) & 0.11\% (11.60s) \\
    tlrmvnmvt (GenzBretz)  & 0.07\% (0.02s) & 0.78\% (0.07s) & 1.02\% (0.14s) & 4.40\% (0.87s) \\
    tlrmvnmvt (TLRQMC)     & 0.05\% (0.02s) & 0.50\% (0.08s) & 1.39\% (0.16s) & 4.43\% (0.78s) \\
    pmnorm (mnorm)         & 0.05\% (0.01s) & 0.14\% (0.05s) & 0.08\% (0.16s) & 2.34\% (2.40s) \\
    \bottomrule
  \end{tabular}
\end{table}

% Table 3
\begin{table}[htbp]
\footnotesize
  \centering
  \caption{Correlation 0.8, upper integration limits -1.}
  \label{tab:3}
  \begin{tabular}{lcccc}
    \toprule
    & 16 & 64 & 128 & 512 \\
    \midrule
    mvtnorm                & 0.02\% (0.04s) & 0.06\% (0.19s) & 0.09\% (0.45s) & 0.12\% (3.36s) \\
    TruncatedNormal        & 0.13\% (0.04s) & 0.14\% (0.26s) & 0.09\% (0.83s) & 0.14\% (10.52s) \\
    tlrmvnmvt (GenzBretz)  & 0.02\% (0.02s) & 0.07\% (0.06s) & 0.13\% (0.14s) & 0.31\% (0.87s) \\
    tlrmvnmvt (TLRQMC)     & 0.03\% (0.02s) & 0.14\% (0.08s) & 0.15\% (0.16s) & 0.27\% (0.78s) \\
    pmnorm (mnorm)         & 0.04\% (0.01s) & 0.01\% (0.06s) & 0.01\% (0.19s) & 0.19\% (2.56s) \\
    \bottomrule
  \end{tabular}
\end{table}

The results are similar (identical in terms of accuracy but slightly differ in terms of speed since we ran the code using another computer) to those reported in \citep{tlrmvnmvt} and suggest that the speed and accuracy of \code{pmnorm} (from \textbf{mnorm}) are comparable to those of the other functions. Therefore, it is reasonable to consider \code{pmnorm} for calculation of the high-dimensional probabilities.

\section{Formulas for the derivatives}\label{sec:derivatives}

In this section, we derive the formulas used in the package to calculate the derivatives of the MVN probabilities and densities. To simplify the derivations, suppose that $\mu$ is a vector of zeros. To apply the formulas from this section to the case when $\mu$ is not a vector of zeros, simply replace $x$ with $x-\mu$. Moreover, for brevity, $f(x;\mu,\Sigma)$, $F(x;\mu,\Sigma)$, and $\overline{F}(x^{(l)}, x^{(u)};\mu,\Sigma)$ are denoted by $f(x)$, $F(x)$, and $\overline{F}(x^{(l)}, x^{(u)})$, respectively.

\subsection{Useful formulas from the linear algebra}

Let $c$, $x$, and $A$ be a constant, a row vector, and a matrix, respectively. A matrix $U=U(c)$ is a function of the constant $c$. We will frequently use the following well-known formulas:

\begin{align}
\frac{\partial x A x^{T}}{\partial x} &= 2xA,\label{eq:b1} \\ 
\frac{\partial \ln \det\left(A\right)}{\partial A} &= \left(A^{-1}\right)^{T}, \label{eq:b2} \\
\frac{\partial xA^{-1}x^{T}}{\partial A} &=-\left(A^{-1}\right)^{T}x^{T}x\left(A^{-1}\right)^{T}, \label{eq:b3} \\
\frac{\partial x A}{\partial x} &= A^{T},\label{eq:b4} \\
\frac{\partial AUB}{\partial c} &= A \frac{\partial U}{\partial c} B,\label{eq:b5} \\
\frac{\partial U^{-1}}{\partial c} &= -U^{-1}\frac{\partial U}{\partial c}U^{-1},\label{eq:b6} \\
\frac{\partial UV}{\partial c} &= U \frac{\partial V}{\partial c} + \frac{\partial U}{\partial c}V.\label{eq:b7} 
\end{align}

Moreover, we use the following lemma.

\begin{lemma}\label{lemma:linal1}
Consider matrices $M$ and $A$. If the product $MA$ is defined and the $i$-th row of the matrix $M$ is a vector of zeros, then the $i$-th row of the matrix $MA$ is also a vector of zeros. Similarly, if the product $AM$ is defined and the $i$-th column of the matrix $M$ is a vector of zeros, then the $i$-th column of the matrix $AM$ is also a vector of zeros.
\end{lemma}
\startproof{of Lemma}{lemma:linal1}
\begin{proofof}
The proof is trivial but we provide it for clarity. Denote the number of rows and columns of the matrix $M$ by $r_{M}$ and $c_{M}$, respectively. Similarly, denote by $r_{A}$ and $c_{A}$ the number of rows and columns of the matrix $A$.

Suppose that $c_{M} = r_{A}$ and the $i$-th row of the matrix $M$ is a vector of zeros, so $M_{ik} = 0$ for any $k\in\{1,...,c_{M}\}$. Hence, the $i$-th row of the matrix $MA$ is also a vector of zeros, because for any $j\in\{1,...,c_{A}\}$, we have:
\begin{equation}
    \left(MA\right)_{ij} = \sum\limits_{k = 1}^{c_{M}} M_{ik}\times A_{kj} = \sum\limits_{k = 1}^{c_{M}}0\times A_{kj} = 0.
\end{equation}
The second claim of the lemma follows from the first one.
\end{proofof}

\subsection{Density function}

In this section, we provide the derivatives of the log-density function of $X$ with respect to the argument $x$ and the elements of the covariance matrix $\Sigma$. The notation $\Sigma_{ij}^{-1}$ corresponds to the element in the $i$-th row and the $j$-th column of the inverse matrix $\Sigma^{-1}$.

% Производная функции плотности
\begin{lemma}\label{lemma:mvn_den_diff}
The derivatives with respect to the argument and the covariance matrix:
\begin{align}
    \nabla_{x}\ln f(x) & = -x\Sigma^{-1}, \label{eq:d1} \\ % аргумент
    \frac{\partial \ln f(x)}{\partial \Sigma} &= \frac{1}{2}\left(\Sigma^{-1}x^{T}x\Sigma^{-1} - \Sigma^{-1}\right) = \nonumber\\
&=\frac{1}{2}\left(\left(\nabla_{x}\ln f(x)\right)^{T} \nabla_{x}\ln f(x) - \Sigma^{-1}\right).\label{eq:d2} % ковариационная матрица
\end{align}
Under the constraint $\Sigma_{ij} = \Sigma_{ji}$, the derivatives with respect to the elements of the covariance matrix are as follows:
\begin{align}
    \frac{\partial \ln f(x)}{\partial \Sigma_{ij}} = \left(\left(\nabla_{x}\ln f(x)\right)_{i} \left(\nabla_{x}\ln f(x)\right)_{j}  - \Sigma^{-1}_{ij}\right) / \left(1 + \mathbb{I}(i = j)\right).\label{eq:d5}
\end{align}
\end{lemma}
\startproof{of Lemma}{lemma:mvn_den_diff}
\begin{proofof}
By using the formula (\ref{eq:b1}), we get the expression (\ref{eq:d1}):
\begin{equation}
\nabla_{x}\ln f(x) = -\frac{1}{2}\frac{\partial x\Sigma^{-1}x^{T}}{\partial x}=-x\Sigma^{-1}.
\end{equation}

By using formulas (\ref{eq:b2}) and (\ref{eq:b3}), we obtain the expression (\ref{eq:d2}):
\begin{align}
\frac{\partial \ln f(x)}{\partial \Sigma} &= -\frac{1}{2} \left(\frac{\partial\ln\det\left(\Sigma\right)}{\partial\Sigma} + \frac{\partial(x)\Sigma^{-1}(x)^{T}}{\partial\Sigma}\right)=\nonumber\\
&=-\frac{1}{2}\left(\Sigma^{-1} - \Sigma^{-1}x^{T}x\Sigma^{-1}\right) = \nonumber\\
&=\frac{1}{2}\left(\left(x\Sigma^{-1}\right)^{T}\left(x\Sigma^{-1}\right) - \Sigma^{-1}\right)=\nonumber\\
&=\frac{1}{2}\left(\left(\nabla_{x}\ln f(x)\right)^{T} \nabla_{x}\ln f(x) - \Sigma^{-1}\right).
\end{align}

Notice that:
\begin{equation}
\left(\left(\nabla_{x}\ln f(x)\right)^{T} \nabla_{x}\ln f(x)\right)_{ij} = \left(\nabla_{x}\ln f(x)\right)_{i} \left(\nabla_{x}\ln f(x)\right)_{j}.
\end{equation}

By using this equality, the formula (\ref{eq:d2}), and addressing the fact that $\Sigma_{ji}$ changes along with $\Sigma_{ij}$, for $i\ne j$, we get:
\begin{align}
    \frac{\partial \ln f(x)}{\partial \Sigma_{ij}} &= \left(\frac{\partial \ln f(x)}{\partial \Sigma}\right)_{ij} + \left(\frac{\partial \ln f(x)}{\partial \Sigma}\right)_{ji} = \nonumber \\
    &= \frac{1}{2}\left(\left(\nabla_{x}\ln f(x)\right)_{i} \left(\nabla_{x}\ln f(x)\right)_{j}  - \Sigma^{-1}_{ij}\right) + \nonumber \\
    &+ \frac{1}{2}\left(\left(\nabla_{x}\ln f(x)\right)_{j} \left(\nabla_{x}\ln f(x)\right)_{i}  - \Sigma^{-1}_{ji}\right) = \nonumber \\
    &= \left(\nabla_{x}\ln f(x)\right)_{i} \left(\nabla_{x}\ln f(x)\right)_{j}  - \Sigma^{-1}_{ij}. \label{eq:d3}
\end{align}

Similarly, the formula (\ref{eq:d2}) implies:
\begin{equation}
\frac{\partial \ln f(x)}{\partial \Sigma_{ii}} = \left(\nabla_{x}\ln f(x)\right)_{i}^2  - \Sigma^{-1}_{ii}.\label{eq:d4}
\end{equation}

Therefore, the expression (\ref{eq:d5}) follows from the formulas (\ref{eq:d3}) and (\ref{eq:d4}).
\end{proofof}

\subsection{Conditional expectation}

In this section, we provide the derivatives of the conditional expectation $\mu_{c}$ with respect to the fixed (given) values $x^{(g)}$ and the covariance matrix $\Sigma$.

% Производные условного математического ожидания
\begin{lemma}\label{lemma:mvn_cmu_diff}
The derivative with respect to the given (fixed) values:
\begin{equation}
\nabla_{x^{(g)}}\mu_{c} = \frac{\partial \mu_{c}}{\partial x^{(g)}} = \left(\Sigma_{d,g}\Sigma^{-1}_{g,g}\right).\label{eq:cm1}
\end{equation}

The derivative with respect to the elements of the covariance matrix such that $i\in I_{d}$ and $j\in I_{g}$:
\begin{gather}
\frac{\partial \mu_{c}}{\partial \Sigma_{ij}} = \left(x^{(g)}-\mu_{g}\right)\left(\Sigma^{-1}_{g,g}M(q_{g}(j), q_{d}(i); m_{g}, m_{d})\right).
\label{eq:cm2}
\end{gather}

The derivative with respect to the elements of the covariance matrix such that $i\in I_{g}$ and $j\in I_{g}$:
\begin{gather}
\frac{\partial \mu_{c}}{\partial \Sigma_{ij}} = \left(x^{(g)}-\mu_{g}\right)\left(-\Sigma^{-1}_{g,g}M(q_{g}(i), q_{g}(j); m_{g})\Sigma^{-1}_{g,g}\Sigma_{g, d}\right).\label{eq:cm3}
\end{gather}

For any $i\in I_{d}$, $j\in I_{g}$, and $k\ne q_{d}(i)$, the following holds:
\begin{gather}
\frac{\partial \mu_{c,k}}{\partial \Sigma_{ij}} = 0, \label{eq:impl_cm2}
\end{gather}
where $\mu_{c,k}$ is the $k$-th element of the vector $\mu_{c}$.
\end{lemma}
\startproof{of Lemma}{lemma:mvn_cmu_diff}
\begin{proofof}
The formula (\ref{eq:cm1}) follows directly from the expression (\ref{eq:b4}). To obtain the formula (\ref{eq:cm2}), transpose $\Sigma_{(d,g)}\Sigma^{-1}_{(g,g)}$ and apply the expression (\ref{eq:b5}). The formula (\ref{eq:cm3}) may be derived similarly to the derivation of the expression (\ref{eq:cm1}), but by applying the expression (\ref{eq:b6}). The formula (\ref{eq:impl_cm2}) follows from the expression (\ref{eq:b5}) and Lemma \ref{lemma:linal1}.
\end{proofof}

\subsection{Conditional covariance}

In this section, we provide the derivatives of the elements of the conditional covariance matrix $\Sigma_{c}$ with respect to the elements of the unconditional covariance matrix $\Sigma$.

% Производные условной ковариационной матрицы
\begin{lemma}\label{lemma:mvn_cs_diff}
The derivatives with respect to the elements of the covariance matrix such that $i\in I_{d}$ and $j\in I_{d}$:
\begin{equation}
\frac{\partial \Sigma_{c, kl}}{\partial \Sigma_{ij}} = 
\begin{cases}
1\text{, if }k = q_{d}(i) \text{ and } l = q_{d}(j)\\
1\text{, if }k = q_{d}(j) \text{ and } l = q_{d}(i)\\
0\text{, otherwise}
\end{cases},\label{eq:cs1}
\end{equation}
where $\Sigma_{c, kl}$ is the element of $\Sigma_{c}$ from the $k$-th row and $l$-th column.

The derivatives with respect to the elements of the covariance matrix such that $i\in I_{d}$ and $j\in I_{g}$:
\begin{align}
\frac{\partial \Sigma_{c}}{\partial \Sigma_{ij}} = &-M(q_{d}(i), q_{g}(j); m_{d}, m_{g})\Sigma^{-1}_{g,g}\Sigma_{g, d} - \nonumber\\ 
&-\Sigma_{d, g}\Sigma^{-1}_{g,g}M(q_{g}(j), q_{d}(i); m_{g}, m_{d}).
\label{eq:cs2}
\end{align}

The derivatives with respect to the elements of the covariance matrix such that $i\in I_{g}$ and $j\in I_{g}$:
\begin{gather}
\frac{\partial \Sigma_{c}}{\partial \Sigma_{ij}} = \Sigma_{d,g}\Sigma_{g,g}^{-1}M(q_{g}(i), q_{g}(j); m_{g})\Sigma_{g,g}^{-1}\Sigma_{g,d}.\label{eq:cs3}
\end{gather}

For any $k,l\in\{1,...,m_{d}\}$, $i\in I_{d}$, and $j\in I_{g}$, the following holds:
\begin{align}
&\frac{\partial \Sigma_{c, kl}}{\partial \Sigma_{ij}} = \nonumber \\ 
&= \left\{
\begin{array}{ll}
-(1 + \mathbb{I}(k = l))\left(\Sigma_{d, g}\Sigma^{-1}_{g,g}\right)_{w_{1}, w_{2}}
  & \text{, if } k = q_{d}(i) \text{ or } l = q_{d}(i) \\[4pt]
0 & \text{, otherwise}
\end{array}
\right.
,\label{eq:impl_cs2}
\end{align}
where $w_{1} = \min(k, l)$ and $w_{2} = \max(k, l)$.
\end{lemma}
\startproof{of Lemma}{lemma:mvn_cs_diff}
\begin{proofof}
The formula (\ref{eq:cs1}) is implied by the expression (\ref{eq:b5}). It is straightforward to derive the formula (\ref{eq:cs2}) by using the expressions (\ref{eq:b5}) and (\ref{eq:b7}). The formula (\ref{eq:cs3}) may be derived from the expressions (\ref{eq:b5}) and (\ref{eq:b6}). The formula (\ref{eq:impl_cs2}) follows from the application of Lemma \ref{lemma:linal1} to both sides of the expression (\ref{eq:cs2}).
\end{proofof}

\subsection{Conditional density function}

In this section, we provide the derivatives of the conditional log-density of $X$ with respect to the argument $x^{(d)}$, the given values $x^{(g)}$, and the elements of the covariance matrix $\Sigma$. For simplicity, suppose that $\mu_{d}$ and $\mu_{g}$ are vectors of zeros. To relax this assumption, simply replace $x_{d}$ and $x_{g}$ with $x_{d}-\mu_{d}$ and $x_{g}-\mu_{g}$, respectively. Finally, for brevity, denote the conditional density function $f\left(x^{(d)};\mu,\Sigma | x^{(g)},I_{g}\right)$ as $f(x^{(d)}\mid x^{(g)})$.

% Производная функции плотности
\begin{lemma}\label{lemma:mvn_cden_diff}
The gradient with respect to the argument:
\begin{gather}
\nabla_{x^{(d)}}\ln f(x^{(d)}\mid x^{(g)}) = -\left(x^{(d)} - \mu_{c}\right)\Sigma_{c}^{-1}. \label{eq:dc1}
\end{gather}

The gradient with respect to the given values:
\begin{gather}
\nabla_{x^{(g)}}\ln f(x^{(d)}\mid x^{(g)}) = -\nabla_{x^{(d)}}\ln f(x^{(d)}\mid x^{(g)})\Sigma_{d,g}\Sigma_{g, g}^{-1}. \label{eq:dc2}
\end{gather}

The derivative with respect to the elements of the covariance matrix such that $i\in I_{d}$ and $j\in I_{d}$:
\begin{align}
\frac{\partial \ln f(x^{(d)}\mid x^{(g)})}{\partial \Sigma_{ij}} &= \left(\frac{\partial \ln f(x^{(d)}\mid x^{(g)})}{\partial x^{(d)}_{q_{d}(i)}}\frac{\partial \ln f(x^{(d)}\mid x^{(g)})}{\partial x^{(d)}_{q_{d}(j)}} - \Sigma_{c,q_{d}(i),q_{d}(j)}^{-1}\right) / \nonumber \\ &/ \left(1 + \mathbb{I}(i = j)\right). \label{eq:dc3}
\end{align}

The derivative with respect to the elements of the covariance matrix such that $i\in I_{d}$ and $j\in I_{g}$:
\begin{gather}
\frac{\partial \ln f(x^{(d)}\mid x^{(g)})}{\partial \Sigma_{ij}} = -\frac{\partial \ln f(x^{(d)}\mid x^{(g)})}{\partial x^{(d)}_{ q_{d}(i)}}\left(\left(x^{(g)} - \mu_{g}\right)\Sigma^{-1}_{g,g}\right)_{q_{g}(j)} - \nonumber \\
-\sum\limits_{k=1}^{m_{d}}\frac{\partial \ln f(x^{(d)}\mid x^{(g)})}{\partial \Sigma_{iq_{d}^{-1}(k)}}(1 + \mathbb{I}(k = q_{d}(i)))\left(\Sigma^{-1}_{g,g}\Sigma_{g, d}\right)_{k, q_{g}(j)}, \label{eq:dc4}
\end{gather}
where the expressions of $\frac{\partial \ln f(x^{(d)}\mid x^{(g)})}{\partial x^{(d)}_{ q_{d}(i)}}$ and $\frac{\partial \ln f(x^{(d)}\mid x^{(g)})}{\partial \Sigma_{iq_{d}^{-1}(k)}}$ are provided in the formulas (\ref{eq:dc1}) and (\ref{eq:dc3}), respectively.

The derivative with respect to the elements of the covariance matrix such that $i\in I_{g}$ and $j\in I_{g}$:
\begin{align}
\frac{\partial \ln f(x^{(d)}\mid x^{(g)})}{\partial \Sigma_{ij}} &= \nabla_{x^{(d)}}\ln f(x^{(d)}\mid x^{(g)})\left(x^{(g)}S\right)^{T} + \nonumber\\
&+ \sum\limits_{k=1}^{m_{d}}\sum\limits_{l=1}^{k}\frac{\partial \ln f(x^{(d)}\mid x^{(g)})}{\partial \Sigma_{q^{-1}_{d}(k)q^{-1}_{d}(l)}}\left(\Sigma_{d,g}S\right)_{q^{-1}_{d}(k), q^{-1}_{d}(l)},\label{eq:dc5}
\end{align}
where:
\begin{gather}
S = \Sigma^{-1}_{g,g}M(q_{g}(i), q_{g}(j); m_{g})\Sigma^{-1}_{g,g}\Sigma_{g,d}.
\end{gather}

The expressions $\nabla_{x^{(d)}}\ln f(x^{(d)}\mid x^{(g)})$ and $\frac{\partial \ln f(x^{(d)}\mid x^{(g)})}{\partial \Sigma_{q_{d}^{-1}(k)q_{d}^{-1}(l)}}$ are provided in the formulas (\ref{eq:dc1}) and (\ref{eq:dc3}), respectively.
\end{lemma}
\startproof{of Lemma}{lemma:mvn_cden_diff}
\begin{proofof}
The formula (\ref{eq:dc1}) may be derived in the same way as the expression (\ref{eq:d1}). 

Further, let's derive the formula (\ref{eq:dc2}). It is fairly straightforward to show that:
\begin{equation}
\nabla_{\mu_{c}}\ln f(x^{(d)}\mid x^{(g)}) = -\nabla_{x^{(d)}}\ln f(x^{(d)}\mid x^{(g)}).
\end{equation}

By using this equality and the formula (\ref{eq:cm1}), we get the expression (\ref{eq:dc2}):
\begin{align}
\nabla_{x^{(g)}}\ln f(x^{(d)}\mid x^{(g)}) &= \frac{\partial \ln f(x^{(d)}\mid x^{(g)})}{\partial \mu_{c}}\frac{\partial \mu_{c}}{\partial x^{(g)}}= \nonumber \\
&=-\nabla_{x^{(d)}}\ln f(x^{(d)}\mid x^{(g)})\Sigma_{d,g}\Sigma_{g, g}^{-1}.
\end{align}

Let's derive the formula (\ref{eq:dc3}). Notice that:
\begin{align}
\frac{\partial \ln f(x^{(d)}\mid x^{(g)})}{\partial \Sigma_{ij}} &= \sum\limits_{k=1}^{m_{d}}\sum\limits_{l=1}^{k}\frac{\partial \ln f(x^{(d)}\mid x^{(g)})}{\partial \Sigma_{c, kl}}\frac{\partial \Sigma_{c, kl}}{\partial \Sigma_{ij}} + \nonumber \\ &+  \sum\limits_{k=1}^{m_{d}}\frac{\partial \ln f(x^{(d)}\mid x^{(g)})}{\partial \mu_{c, k}}\frac{\partial \mu_{c, k}}{\partial \Sigma_{ij}}.
\end{align}

Since $\mu_{c}$ is independent of $\Sigma_{d,d}$, the second summand in the last formula is zero. By using the formula (\ref{eq:cs1}), we get $\frac{\partial \Sigma_{c,kl}}{\partial \Sigma_{ij}}=1$ if and only if 1) $q_{d}(i) = k$ and $q_{d}(j) = l$ or 2) $q_{d}(j) = k$ and $q_{d}(i) = l$. Otherwise, the derivative is $0$. Hence, by applying the expression (\ref{eq:d5}) to the non-zero terms, we get the formula (\ref{eq:dc3}):
\begin{gather}
\sum\limits_{k=1}^{m_{d}}\sum\limits_{l=1}^{k}\frac{\partial \ln f(x^{(d)}\mid x^{(g)})}{\partial \Sigma_{c, kl}}\frac{\partial \Sigma_{c, kl}}{\partial \Sigma_{ij}} = \frac{\partial \ln f(x^{(d)}\mid x^{(g)})}{\partial \Sigma_{c, q_{d}(i)q_{d}(j)}}= \nonumber \\
=\left(\frac{\partial \ln f(x^{(d)}\mid x^{(g)})}{\partial x^{(d)}_{q_{d}(i)}}\frac{\partial \ln f(x^{(d)}\mid x^{(g)})}{\partial x^{(d)}_{q_{d}(j)}} - \Sigma_{c,q_{d}(i)q_{d}(j)}^{-1}\right) / \left(1 + \mathbb{I}(i = j)\right).
\end{gather}

Let's derive the formula (\ref{eq:dc4}). Notice that:
\begin{align}
\frac{\partial \ln f(x^{(d)}\mid x^{(g)})}{\partial \Sigma_{ij}} &=\sum\limits_{k=1}^{m_{d}}\sum\limits_{l=1}^{k}\frac{\partial \ln f(x^{(d)}\mid x^{(g)})}{\partial \Sigma_{c, kl}}\frac{\partial \Sigma_{c, kl}}{\partial \Sigma_{ij}} + \nonumber \\&+ \sum\limits_{k=1}^{m_{d}}\frac{\partial \ln f(x^{(d)}\mid x^{(g)})}{\partial \mu_{c, k}}\frac{\partial \mu_{c, k}}{\partial \Sigma_{ij}}.
\end{align}

Consider the left-hand and right-hand sides of the last expression. By applying the formula (\ref{eq:cs2}) to the left-hand side, we get:
\begin{gather}
\sum\limits_{k=1}^{m_{d}}\sum\limits_{l=1}^{k}\frac{\partial \ln f(x^{(d)}\mid x^{(g)})}{\partial \Sigma_{c, kl}}\frac{\partial \Sigma_{c, kl}}{\partial \Sigma_{ij}} = \sum\limits_{k=1}^{m_{d}} \frac{\partial \ln f(x^{(d)}\mid x^{(g)})}{\partial \Sigma_{c, iq_{d}^{-1}(k)}}\frac{\partial \Sigma_{c, iq_{d}^{-1}(k)}}{\partial \Sigma_{ij}} = \nonumber \\
= \sum\limits_{k=1}^{m_{d}}\frac{\partial \ln f(x^{(d)}\mid x^{(g)})}{\partial \Sigma_{iq_{d}^{-1}(k)}}(1 + \mathbb{I}(k = q_{d}(i)))\left(-\Sigma_{d, g}\Sigma^{-1}_{g,g}\right)_{k,q_{g}(j)},
\end{gather}
where the substitution of $\Sigma_{c,kl}$ with $\Sigma_{iq^{-1}_{d}(k)}$ in the denominator is justified by the formula (\ref{eq:cs1}). By applying the expression (\ref{eq:cm2}) to the right-hand side and combining it with the left-hand side, we obtain the formula (\ref{eq:dc4}).

Finally, let's derive the formula (\ref{eq:dc5}). Notice that:
\begin{align}
\frac{\partial \ln f(x^{(d)}\mid x^{(g)})}{\partial \Sigma_{ij}} = &\sum\limits_{k=1}^{m_{d}}\sum\limits_{l=1}^{k}\frac{\partial \ln f(x^{(d)}\mid x^{(g)})}{\partial \Sigma_{c,kl}}\frac{\partial \Sigma_{c,kl}}{\partial \Sigma_{ij}} + \nonumber \\ &\sum\limits_{k=1}^{m_{d}}\frac{\partial \ln f(x^{(d)}\mid x^{(g)})}{\partial \mu_{c, k}}\frac{\partial \mu_{c, k}}{\partial \Sigma_{ij}}.
\end{align}

Consider the left-hand and right-hand sides of the last expression. By applying the formula (\ref{eq:cs3}) to the left-hand side, we get:
\begin{align}
\sum\limits_{k=1}^{m_{d}}\sum\limits_{l=1}^{k}&\frac{\partial \ln f(x^{(d)}\mid x^{(g)})}{\partial \Sigma_{c,kl}}\frac{\partial \Sigma_{c,kl}}{\partial \Sigma_{ij}} 
= \sum\limits_{k=1}^{m_{d}}\sum\limits_{l=1}^{k}\frac{\partial \ln f(x^{(d)}\mid x^{(g)})}{\partial \Sigma_{q^{-1}_{d}(k), q^{-1}_{d}(l)}}\times \nonumber \\ &\times \left(\Sigma_{d,g}\Sigma_{g,g}^{-1}M(q_{g}(i),q_{g}(j); m_{g})\Sigma_{g,g}^{-1}\Sigma_{g,d}\right)_{q^{-1}_{d}(k), q^{-1}_{d}(l)},
\end{align}
where the replacement of $\Sigma_{c,kl}$ with $\Sigma_{q^{-1}_{d}(k)q^{-1}_{d}(l)}$ in the denominator is justified by the formula (\ref{eq:cs1}).

By applying the expression (\ref{eq:cm3}) to the right-hand side and rewriting the sum in a matrix form, we get:
\begin{align}
\sum\limits_{k=1}^{m_{d}}\frac{\partial \ln f(x^{(d)}\mid x^{(g)})}{\partial \mu_{c, k}}\frac{\partial \mu_{c, k}}{\partial \Sigma_{ij}} 
&=\nabla_{x^{(d)}}\ln f(x^{(d)}\mid x^{(g)}) \times \nonumber \\ &\times \left(x^{(g)}\left(\Sigma^{-1}_{g,g}M(q_{g}(i), q_{g}(j); m_{g})\Sigma^{-1}_{g,g}\Sigma_{g,d}\right)\right)^{T}.
\end{align}

By combining the obtained expressions of the left-hand and right-hand sides, we get the formula (\ref{eq:dc5}).
\end{proofof}

\subsection{Cumulative distribution function}

In this section, we provide the derivatives of the cumulative distribution function of $X$ with respect to the argument $x$ and the elements of the covariance matrix $\Sigma$.

% Производная функции распределения
\begin{lemma}\label{lemma:mvn_cdf_diff}
The derivative with respect to the argument:
\begin{equation}
\frac{\partial F\left(x\right)}{\partial x_{i}} = \mathbb{P}\left(X_{(-i)}\leq x_{(-i)}|X_{i}=x_{i}\right)f_{X_{i}}\left(x_{i}\right)\label{eq:p1}.
\end{equation}

The derivative with respect to the covariances for $i\ne j$:
\begin{align}
\frac{\partial F\left(x\right)}{\partial \Sigma_{ij}} &= \frac{\partial^2 F\left(x\right)}{\partial x_{i} \partial x_{j}} = \nonumber \\ &= \mathbb{P}\left(X_{(-i,-j)}\leq x_{(-i,-j)}|X_{i}=x_{i},X_{j}=x_{j}\right)f_{X_{i},X_{j}}\left(x_{i}, x_{j}\right)\label{eq:p2}.
\end{align}

The derivative with respect to the variance:
\begin{equation}
\frac{\partial F\left(x\right)}{\partial \Sigma_{ii}} = -\frac{1}{2\Sigma_{ii}}\left(\frac{\partial F(x)}{\partial x_{i}}x_{i} + \sum\limits_{j\ne i}\frac{\partial F(x)}{\partial \Sigma_{ij}}\Sigma_{ij}\right),\label{eq:p3}
\end{equation}
where $\frac{\partial F(x)}{\partial x_{i}}$ and $\frac{\partial F(x)}{\partial \Sigma_{ij}}$ may be calculated by the formulas (\ref{eq:p1}) and (\ref{eq:p2}), respectively.

\end{lemma}
\startproof{of Lemma}{lemma:mvn_cdf_diff}
\begin{proofof}
Without loss of generality, suppose that $i = 1$. 

The formula (\ref{eq:p1}) is derived by the following steps:
\begin{align}
\frac{\partial F\left(x\right)}{\partial x_{1}} &= \frac{\partial \int\limits_{-\infty}^{x_{1}}...\int\limits_{-\infty}^{x_{m}} f(x_{1},...,x_{m})dx_{m}...dx_{1}}{\partial x_{1}} = \nonumber \\
&= \int\limits_{-\infty}^{x_{2}}...\int\limits_{-\infty}^{x_{m}} f(x_{1},...,x_{m})dx_{m}...dx_{2} = \nonumber\\
&= \int\limits_{-\infty}^{x_{2}}...\int\limits_{-\infty}^{x_{m}} f_{X_{(-1)}|X_{1}=x_{1}}(x_{2},...,x_{m})f_{X_{1}}(x_{1})dx_{m}...dx_{2}= \nonumber\\
&=f_{X_{1}}(x_{1})\int\limits_{-\infty}^{x_{2}}...\int\limits_{-\infty}^{x_{m}} f_{X_{(-1)}|X_{1}=x_{1}}(x_{2},...,x_{m})dx_{m}...dx_{2}= \nonumber\\
&=f_{X_{1}}(x_{1})\mathbb{P}\left(X_{(-1)}\leq x_{(-1)}|X_{1}=x_{1}\right).
\end{align}

Next, we derive the formula (\ref{eq:p2}). By using the expression (\ref{eq:d3}), it is fairly straightforward to show that:
\begin{equation}
\frac{\partial f(x)}{\partial \Sigma_{ij}} = \frac{\partial^2 f(x)}{\partial x_{i}\partial x_{j}}.
\end{equation}

By using the last equality and (without loss of generality) setting $j=2$, we get the formula (\ref{eq:p2}):
\begin{align}
\frac{\partial F\left(x\right)}{\partial \Sigma_{12}} &= \frac{\partial^2 F\left(x\right)}{\partial x_{1} \partial x_{2}} = \frac{\partial \int\limits_{-\infty}^{x_{1}}...\int\limits_{-\infty}^{x_{m}} f(x_{1},...,x_{m})dx_{m}...dx_{1}}{\partial x_{1} \partial x_{2}} = \nonumber \\
&= \int\limits_{-\infty}^{x_{3}}...\int\limits_{-\infty}^{x_{m}} f(x_{1},...,x_{m})dx_{m}...dx_{3} = \nonumber \\
&= \int\limits_{-\infty}^{x_{3}}...\int\limits_{-\infty}^{x_{m}} f_{X_{(-1, -2)}|X_{1}=x_{1},X_{2} = x_{2}}(x_{3},...,x_{m})\times \nonumber \\ & \times f_{X_{1},X_{2}}(x_{1},x_{2})dx_{m}...dx_{3}= \nonumber \\
&=f_{X_{1},X_{2}}(x_{1},x_{2}) \times \nonumber \\ &\times \int\limits_{-\infty}^{x_{3}}...\int\limits_{-\infty}^{x_{m}} f_{X_{(-1,-2)}|X_{1}=x_{1},X_{2}=x_{2}}(x_{3},...,x_{m})dx_{m}...dx_{3}= \nonumber \\
&=f_{X_{1},X_{2}}\left(x_{1}, x_{2}\right)\mathbb{P}\left(X_{(-1,-2)}\leq x_{(-1,-2)}|X_{1}=x_{1},X_{2}=x_{2}\right).
\end{align}

Finally, we derive the formula (\ref{eq:p3}). To do so, we introduce a new vector:
\begin{equation}
\tilde{X} = \left(\frac{X_{1}}{\sqrt{\Sigma_{1, 1}}},X_{2},...,X_{m}\right).
\end{equation}

Notice that $\tilde{X}$ is similar to $X$. Specifically, the only difference is that $\tilde{X}_{1}$ is actually a standardized (to unit variance) version of $X_{1}$. Therefore, the covariance matrix of $\tilde{X}$ is as follows:

\begin{equation}
\text{Cov}\left(\tilde{X}\right)=\tilde{\Sigma}=
\begin{bmatrix}
1 & \frac{\Sigma_{12}}{\sqrt{\Sigma_{11}}} & \frac{\Sigma_{13}}{\sqrt{\Sigma_{11}}} & ... & \frac{\Sigma_{1m}}{\sqrt{\Sigma_{11}}}\\
\frac{\Sigma_{12}}{\sqrt{\Sigma_{11}}} & \Sigma_{22} & \Sigma_{23} & ... & \Sigma_{2m}\\
\frac{\Sigma_{13}}{\sqrt{\Sigma_{11}}} & \Sigma_{23} & \Sigma_{33} & ... & \Sigma_{3m}\\
\vdots & \vdots & \vdots & \ddots & \vdots\\
\frac{\Sigma_{1m}}{\sqrt{\Sigma_{11}}} & \Sigma_{2m} & \Sigma_{3m} & ... & \Sigma_{mm}\\
\end{bmatrix}.
\end{equation}

Denote by $\tilde{F}$ the cumulative distribution function of $\tilde{X}$. Notice that it has the following relationship with the cumulative distribution function of $X$:
\begin{equation}
F(x) = \mathbb{P}\left(\frac{X_{1}}{\sqrt{\Sigma_{11}}}\leq \frac{x_{1}}{\sqrt{\Sigma_{11}}},X_{2}\leq x_{2},...,X_{m}\leq x_{m}\right)=\tilde{F}(\tilde{x}),
\end{equation}
where $\tilde{x} = \left(\frac{x_{1}}{\sqrt{\Sigma_{11}}},x_{2},...,x_{m}\right)$.

Finally, by using the last equality, we derive the formula (\ref{eq:p3}):
\begin{align}
\frac{\partial F\left(x\right)}{\partial \Sigma_{11}} &= 
\frac{\partial \tilde{F}\left(\tilde{x}\right)}{\partial \Sigma_{11}} =
\frac{\partial \tilde{F}\left(\tilde{x}\right)}{\partial \tilde{x}_{1}}\frac{\partial \tilde{x}_{1}}{\partial \Sigma_{11}} + \sum\limits_{j=2}^{m} \frac{\partial \tilde{F}\left(\tilde{x}\right)}{\partial \tilde{\Sigma}_{1j}}\frac{\partial \tilde{\Sigma}_{1j}}{\partial \Sigma_{11}}=\nonumber\\
&=-\left(\sqrt{\Sigma_{11}}\frac{\partial F\left(x\right)}{\partial x_{1}}\right)\frac{x_{1}}{2\Sigma_{11}^{1.5}} - \sum\limits_{j=2}^{m} \left(\sqrt{\Sigma_{11}}\frac{\partial F\left(x\right)}{\partial \Sigma_{1j}}\right)\frac{\Sigma_{1j}}{2\Sigma_{11}^{1.5}} = \nonumber\\
&=-\frac{1}{2\Sigma_{11}}\left(\frac{\partial F(x)}{\partial x_{1}}x_{1} + \sum\limits_{j\ne 1}\frac{\partial F(x)}{\partial \Sigma_{1j}}\Sigma_{1j}\right).
\end{align}

\end{proofof}

\subsection{Probabilities}

In this section, we provide the derivatives of the probability of the event that $X$ takes values between $x^{(l)}$ and $x^{(u)}$, with respect to these arguments and the elements of the covariance matrix $\Sigma$.

% Производные вероятностей
\begin{lemma}\label{lemma:mvn_idf_diff}
The derivatives with respect to the upper integration limits:
\begin{align}
\frac{\partial \overline{F}\left(x^{(l)}, x^{(u)}\right)}{\partial x^{(u)}_{i}} &= \frac{\partial \mathbb{P}\left(x^{(l)}\leq X\leq x^{(u)}\right)}{\partial x^{(u)}_{i}} = \nonumber \\ &= \mathbb{P}\left(x^{(l)}_{(-i)}\leq X_{(-i)}\leq x^{(u)}_{(-i)}|X_{i}=x^{(u)}_{i}\right)f_{X_{i}}\left(x^{(u)}_{i}\right).\label{eq:pp1}
\end{align}

The derivatives with respect to the lower integration limits:
\begin{align}
\frac{\partial \overline{F}\left(x^{(l)}, x^{(u)}\right)}{\partial x^{(l)}_{i}} &= \frac{\partial \mathbb{P}\left(x^{(l)}\leq X\leq x^{(u)}\right)}{\partial x^{(l)}_{i}} = \nonumber \\ &= -\mathbb{P}\left(x^{(l)}_{(-i)}\leq X_{(-i)}\leq x^{(u)}_{(-i)}|X_{i}=x^{(l)}_{i}\right)f_{X_{i}}\left(x^{(l)}_{i}\right).\label{eq:pp2}
\end{align}

The derivatives with respect to the covariances for $i\ne j$:
\begin{align}
\frac{\partial  \overline{F}\left(x^{(l)}, x^{(u)}\right)}{\partial \Sigma_{ij}} &= \frac{\partial \mathbb{P}\left(x^{(l)}\leq X\leq 
x^{(u)}\right)}{\partial \Sigma_{ij}} \nonumber \\ &=p_{uu}d_{uu}-p_{lu}d_{lu}-p_{ul}d_{ul}+p_{ll}d_{ll},\label{eq:pp3}
\end{align}
where for $a,b\in\{u,l\}$, the following notations are used:
\begin{align}
p_{ab} &= \mathbb{P}\left(x^{(l)}_{(-(i, j))}\leq X_{-(i, j)}\leq x^{(u)}_{(-(i, j))}|
X_{i}=x^{(a)}_{i}, X_{j}=x^{(b)}_{j}\right), \\
d_{ab} &= f_{X_{i}, X_{j}}\left(x^{(a)}_{i}, x^{(b)}_{j};
\mu_{(i,j)},\Sigma_{(i, j),(i, j)}\right).
\end{align}

The derivative with respect to the variance:
\begin{align}
\frac{\partial \mathbb{P}\left(x^{(l)}\leq X\leq 
x^{(u)}\right)}{\partial \Sigma_{ii}} =-\frac{1}{2\Sigma_{ii}}\left(\frac{\partial P}{\partial x_{i}^{(u)}}x^{(u)}_{i} + \frac{\partial P}{\partial x_{i}^{(l)}}x^{(l)}_{i} + \sum\limits_{j\ne i}\frac{\partial P}{\partial \Sigma_{ij}}\Sigma_{ij}\right),\label{eq:pp4}
\end{align}
where notation $P = \mathbb{P}\left(x^{(l)}\leq X\leq x^{(u)}\right)$ is used for brevity.
\end{lemma}
\startproof{of Lemma}{lemma:mvn_idf_diff}
\begin{proofof}

The formulas (\ref{eq:pp1}), (\ref{eq:pp2}), and (\ref{eq:pp4}) may be derived by the same steps used to obtain the expressions (\ref{eq:p1}) and (\ref{eq:p2}). The derivation of the formula (\ref{eq:pp3}) is trickier, so we consider it in detail.

Without loss of generality, suppose that $i = 1$ and $j = 2$. Notice that:
\begin{gather}
\mathbb{P}\left(x^{(l)}\leq X\leq 
x^{(u)}\right) = \nonumber \\
=\mathbb{P}\left(X_{1}\leq 
x^{(u)}_{1}, X_{2}\leq 
x^{(u)}_{2},x_{3}^{(l)}\leq X_{3}\leq 
x^{(u)}_{3},...,x_{m}^{(l)}\leq X_{m}\leq 
x^{(u)}_{m}\right) - \nonumber\\
-\mathbb{P}\left(X_{1}\leq 
x^{(u)}_{1}, X_{2}\leq x^{(l)}_{2} 
,x_{3}^{(l)}\leq X_{3}\leq 
x^{(u)}_{3},...,x_{m}^{(l)}\leq X_{m}\leq 
x^{(u)}_{m}\right) - \nonumber\\
-\mathbb{P}\left( X_{1}\leq x^{(l)}_{1} 
, X_{2}\leq x^{(u)}_{2}
,x_{3}^{(l)}\leq X_{3}\leq 
x^{(u)}_{3},...,x_{m}^{(l)}\leq X_{m}\leq 
x^{(u)}_{m}\right)+ \nonumber \\
+\mathbb{P}\left( X_{1}\leq x^{(l)}_{1} 
, X_{2} \leq x^{(l)}_{2},x_{3}^{(l)}\leq X_{3}\leq 
x^{(u)}_{3},...,x_{m}^{(l)}\leq X_{m}\leq 
x^{(u)}_{m}\right).
\end{gather}

By applying formula (\ref{eq:p2}) to each of these probabilities, we get the expression (\ref{eq:pp3}).
\end{proofof}

\subsection{Conditional probabilities}

In this section, we provide the derivatives of the conditional probabilities with respect to the integration limits $x^{(l)}$, $x^{(u)}$, and the elements of the covariance matrix $\Sigma$. 

Specifically, we redefine $x^{(l)}$ and $x^{(u)}$ as the integration limits associated only with the components indexed by $I_{d}$. Moreover, for brevity, we introduce the following notations:
\begin{align}
\text{P} &= \mathbb{P}\left(x^{(l)}\leq X^{(d)}\leq x^{(u)}|X^{(g)} = x^{(g)}\right), \\
\tilde{\text{P}} &= \mathbb{P}\left(\tilde{x}^{(l)} \leq \tilde{X} \leq \tilde{x}^{(u)}\right), \\
\tilde{x}^{(l)} &= x^{(l)}-\mu_{c}, \\ 
\tilde{x}^{(u)} &= x^{(u)}-\mu_{c}, \\
\tilde{x}^{(g)}&=x^{(g)}-\mu_{g}, \\
\tilde{X}& \sim\text{N}\left(\left(0,...,0\right), \Sigma_{c}\right).
\end{align}

Despite $\tilde{\text{P}}=\text{P}$, sometimes it is more convenient to differentiate $\tilde{\text{P}}$ rather than $\text{P}$.

% Производные условных вероятностей
\begin{lemma}\label{lemma:mvn_cidf_diff}
The derivative with respect to the upper integration limits, where $i\in\{1,...,m_{d}\}$:
\begin{align}
\frac{\partial \text{P}}{\partial x^{(u)}_{i}} &= \mathbb{P}\left(x^{(l)}_{(-i)}\leq X_{(-i)}^{(d)}\leq x^{(u)}_{(-i)}|X_{i}^{(d)}=x^{(u)}_{i}, X^{(g)}=x^{(g)}\right)\times \nonumber\\ &\times f_{X_{i}^{(d)}|X^{(g)}=x^{(g)}}\left(x^{(u)}_{i}\right)= \nonumber \\
&=\mathbb{P}\left(\tilde{x}^{(l)}_{(-i)}\leq \tilde{X}_{(-i)}^{(d)}\leq \tilde{x}^{(u)}_{(-i)}|\tilde{X}_{i}^{(d)}=\tilde{x}^{(u)}_{i}\right)f_{\tilde{X}_{i}^{(d)}}\left(\tilde{x}^{(u)}_{i}\right).\label{eq:pc1}
\end{align}

The derivative with respect to the lower integration limits, where $i\in\{1,...,m_{d}\}$:
\begin{align}
\frac{\partial \text{P}}{\partial x^{(l)}_{i}} &= -\mathbb{P}\left(x^{(l)}_{(-i)}\leq X_{(-i)}^{(d)}\leq x^{(u)}_{(-i)}|X_{i}^{(d)}=x^{(l)}_{i}, X^{(g)}=x^{(g)}\right)\times\nonumber \\ &\times f_{X_{i}^{(d)}|X^{(g)}=x^{(g)}}\left(x^{(l)}_{i}\right)= \nonumber \\
&= -\mathbb{P}\left(\tilde{x}^{(l)}_{(-i)}\leq \tilde{X}_{(-i)}^{(d)}\leq \tilde{x}^{(u)}_{(-i)}|\tilde{X}_{i}^{(d)}=\tilde{x}^{(l)}_{i}\right)f_{\tilde{X}_{i}^{(d)}}\left(\tilde{x}^{(l)}_{i}\right).\label{eq:pc2}
\end{align}

The derivative with respect to the fixed (conditioned) values:
\begin{align}
\frac{\partial \text{P}}{\partial x^{(g)}_{i}} =-\left(\nabla_{x^{(l)}}\text{P}+\nabla_{x^{(u)}}\text{P}\right)\Sigma_{d,g}\Sigma_{g, g}^{-1},\label{eq:pc3}
\end{align}
where the gradients $\nabla_{x^{(u)}}\text{P}$ and $\nabla_{x^{(l)}}\text{P}$ may be calculated by the formulas (\ref{eq:pc1}) and (\ref{eq:pc2}), respectively.

The derivatives with respect to the covariances such that $i\in I_{d}$, $j\in I_{d}$, and $i\ne j$:
\begin{align}
\frac{\partial \text{P}}{\partial \Sigma_{ij}} = p_{uu}d_{uu}-p_{lu}d_{lu}-p_{ul}d_{ul}+p_{ll}d_{ll}, \nonumber \label{eq:pc4}
\end{align}
where for $a,b\in\{u,l\}$, the following notations are used:
\begin{align}
p_{ab} &= \mathbb{P}\left(x^{(l)}_{(-(i, j))}\leq X_{-(i, j)}^{(d)}\leq x^{(u)}_{(-(i, j))}|
X_{i}=x^{(a)}_{i}, X_{j}=x^{(b)}_{j}, X^{(g)}=x^{(g)}\right), \nonumber \\
d_{ab} &= f_{X_{i}, X_{j}|X^{(g)}=x^{(g)}}\left(x^{(a)}_{i}, x^{(b)}_{j}\right).
\end{align}

The derivative with respect to the variance for $i\in I_{d}$:
\begin{align}
\frac{\partial \text{P}}{\partial \Sigma_{ii}} = -\frac{1}{2\Sigma_{c, q_{d}(i)q_{d}(i)}}\left(\begin{aligned}&\frac{\partial \tilde{\text{P}}}{\partial \tilde{x}_{i}^{(u)}}\tilde{x}^{(u)}_{q_{d}(i)} + \frac{\partial \tilde{\text{P}}}{\partial \tilde{x}_{i}^{(l)}}\tilde{x}^{(l)}_{q_{d}(i)} + \\ &+ \sum\limits_{j\in I_{d}:j\ne i}\frac{\partial \tilde{\text{P}}}{\partial \Sigma_{c, q_{d}(i)q_{d}(j)}}\Sigma_{c, q_{d}(i)q_{d}(j)}\end{aligned}\right).\label{eq:pc5}
\end{align}

The derivative with respect to the covariances such that $i\in I_{d}$ and $j\in I_{g}$:
\begin{align}
\frac{\partial \text{P}}{\partial \Sigma_{ij}} = &-\left(\frac{\partial \text{P}}{\partial x^{(u)}_{q_{d}(i)}}+\frac{\partial \text{P}}{\partial x^{(l)}_{q_{d}(i)}}\right)\left(\left(x^{(g)} - \mu_{g}\right)\Sigma^{-1}_{g,g}\right)_{q_{g}(j)} - \nonumber \\
&-\sum\limits_{k=1}^{m_{d}}\frac{\partial \text{P}}{\partial \Sigma_{iq_{d}^{-1}(k)}}(1 + \mathbb{I}(k = q_{d}(i)))\left(\Sigma^{-1}_{g,g}\Sigma_{g, d}\right)_{k, q_{g}(j)}\label{eq:pc6}.
\end{align}

The derivative with respect to the covariances such that $i\in I_{g}$ and $j\in I_{g}$:
\begin{align}
\frac{\partial \text{P}}{\partial \Sigma_{ij}} &= \left(\nabla_{x^{(l)}}\text{P} + \nabla_{x^{(u)}}\text{P} \right)\left(x^{(g)}S\right)^{T} + \nonumber \\
&+ \sum\limits_{k=1}^{m_{d}}\sum\limits_{l=1}^{k}\frac{\partial \text{P}}{\partial \Sigma_{q^{-1}_{d}(k), q^{-1}_{d}(l)}}\left(\Sigma_{d,g}S\right)_{q^{-1}_{d}(k), q^{-1}_{d}(l)}\label{eq:pc7},
\end{align}
where:
\begin{gather}
S = \Sigma^{-1}_{g,g}M(q_{g}(i), q_{g}(j); m_{g})\Sigma^{-1}_{g,g}\Sigma_{g,d}.
\end{gather}

\end{lemma}
\startproof{of Lemma}{lemma:mvn_cidf_diff}
\begin{proofof}

Notice that $\frac{\partial \text{P}}{\partial x^{(u)}_{i}} = \frac{\partial \tilde{\text{P}}}{\partial \tilde{x}^{(u)}_{i}}$, so the formula (\ref{eq:pc1}) may be derived by the same steps used to obtain the expression (\ref{eq:pp1}). The derivation of the formula (\ref{eq:pc2}) is very similar. The derivation of the formula (\ref{eq:pc3}) is very similar to the derivation of the expression (\ref{eq:dc2}). The formula (\ref{eq:pc4}) may be derived by the same steps used to obtain the expressions (\ref{eq:dc3}) and (\ref{eq:pp3}). The formula (\ref{eq:pc5}) may be derived by the same steps used to obtain the expression (\ref{eq:pp4}). The formula (\ref{eq:pc6}) may be derived by the same steps used to obtain the expression (\ref{eq:dc4}). Finally, the formula (\ref{eq:pc7}) may be derived by the same steps used to obtain the expression (\ref{eq:dc5}).
\end{proofof}

\section{Bounds}\label{sec:bounds}

In this section, we consider some useful bounds on the probabilities of the multivariate normal distribution and their derivatives. In particular, these inequalities are useful for proving consistency and asymptotic normality of maximum-likelihood estimators in econometric models such as the multivariate probit model. Therefore, these bounds will be useful for the users of the package.

\subsection{Probabilities}

In this section, we provide a lower bound on the logarithm of the cumulative distribution function of the multivariate normal distribution.

% Ограничение снизу вероятностей
\begin{lemma}\label{lemma:mvn_cdf_lowerbound}
The following inequality holds:
\begin{align}
    \ln F(x) \geq -\left(x - \mu\right)\Sigma^{-1}\left(x - \mu\right)^{T} - C,
\end{align}
where:
\begin{align}
    C = \frac{m}{2}\ln \left(2\pi\right) + \frac{1}{2}\ln\left(\text{det}\left(\Sigma\right)\right)+||\Sigma^{-1}||_{2} + \ln\left(m!\right).
\end{align}
\end{lemma}
\startproof{of Lemma}{lemma:mvn_cdf_lowerbound}
\begin{proofof}

For brevity, suppose that $\mu = 0$ and consider the $m$-variate row vector $t$, such that:
\begin{align}
    \forall j\in\{1,...,m\}: t_{j} \leq x_{j},\\
    \sum \limits_{j=1}^{m} \left(x_{j} - t_{j}\right) \leq 1.
\end{align}

Denote $\triangle = x - t$ and notice that:
\begin{align}
    \left(x - \triangle\right)\Sigma^{-1}\left(x - \triangle\right)^{T}\leq \nonumber \\ 
    \leq \underbrace{2x\Sigma^{-1}x^{T} + 2\triangle\Sigma^{-1}\triangle^{T}}_{\text{Since }\Sigma^{-1}\text{ is positive-definite}} \leq \nonumber \\
    \underbrace{\leq 2x\Sigma^{-1}x^{T} + 2||\Sigma^{-1}||_{2}||\triangle||^{2}}_{\text{Properties of spectral norm}} \leq  \nonumber \\
    \leq \underbrace{2x\Sigma^{-1}x^{T} + 2||\Sigma^{-1}||_{2}}_{\text{Since }\left(\sum\limits_{j=1}^{m}\triangle_{j}\right)\in\left[0, 1\right]}.
\end{align}

Since the density function $f(x)$ is strictly decreasing with respect to the expression in the exponent, we get:
\begin{equation}\label{eq:mvn_cdf_lowerbound_key}
    f(x - \triangle) \geq \frac{1}{\left(2\pi\right)^{m/2}\sqrt{\text{det}\left(\Sigma\right)}} e^{-x\Sigma^{-1}x^{T} - ||\Sigma^{-1}||_{2}}.
\end{equation}

By using this inequality, we obtain:
\begin{align}
    F(x) &= \int\limits_{-\infty}^{x} f(u) du \geq \int\limits_{-\infty}^{x} \mathbb{I}\left(\sum\limits_{j=1}^{m} \left(x_{j} - u_{j}\right)\leq 1\right)f(u) du= \nonumber \\
    &= \underbrace{\int\limits_{0}^{1}\int\limits_{0}^{1 - u_{1}}...\int\limits_{0}^{1 - u_{1} - ... - u_{m}} f(x - u) du}_{\substack{\text{For }u\text{ the same properties hold}\\ \text{as for }\triangle}} \geq \nonumber \\
    &\geq \underbrace{\int\limits_{0}^{1}\int\limits_{0}^{1 - u_{1}}...\int\limits_{0}^{1 - u_{1} - ... - u_{m}} \frac{1}{\left(2\pi\right)^{m/2}\sqrt{\text{det}\left(\Sigma\right)}} e^{-x\Sigma^{-1}x^{T} - ||\Sigma^{-1}||_{2}} du}_{\text{Inequality }(\ref{eq:mvn_cdf_lowerbound_key})} = \nonumber \\
    &= \frac{1}{\left(2\pi\right)^{m/2}\sqrt{\text{det}\left(\Sigma\right)}} e^{-x\Sigma^{-1}x^{T} - ||\Sigma^{-1}||_{2}} \underbrace{\int\limits_{0}^{1}\int\limits_{0}^{1 - u_{1}}...\int\limits_{0}^{1 - u_{1} - ... - u_{m}} 1 du}_{1/m!} = \nonumber \\
    &= \frac{1}{m!\left(2\pi\right)^{m/2}\sqrt{\text{det}\left(\Sigma\right)}} e^{-x\Sigma^{-1}x^{T} - ||\Sigma^{-1}||_{2}}.
\end{align}
The inequality of the lemma follows by taking the logarithm of both sides of the last inequality.
\end{proofof}

Notice that the lower bound of Lemma \ref{lemma:mvn_cdf_lowerbound} is less strict than the one obtained by \citep{Hashorva2003}. However, the derived inequality may be useful in some practical scenarios due to its simplicity.

\subsection{Derivatives of log-probabilities}

In this section, we provide an upper bound on the derivatives of the log-probabilities of the multivariate normal distribution.

% Ограничение сверху производных вероятностей
\begin{lemma}\label{lemma:mvn_cdf_grad_upperbound}
The following inequality holds:
\begin{align}
    \left\|\nabla_{x} \ln F(x)\right\|_{2} \leq C_{m}\left(\Sigma\right)\left(1 + \left\|x - \mu\right\|_{2}\right),
\end{align}
where $C_{m}\left(\Sigma\right)$ is a function such that:
\begin{align}
    C_{m}\left(\Sigma\right) &= \sum\limits_{i=1}^{m} c_{i}, \nonumber \\
    c_{i} &= 
    \begin{cases}
    \max\left(\frac{1}{\Sigma_{ii}}, \frac{1}{\sqrt{\Sigma_{ii}}}\right)\text{, if }m=1\text{ or }\forall j\ne i:\Sigma_{ij} = 0\\
    \frac{2}{M_{i}^{*}}\left(1+\frac{M_{i}}{\ln\left(2\right)}\right)\text{, otherwise}
    \end{cases}, \\
     M_{i} &= \left\|\frac{\Sigma_{-i,i}}{\Sigma_{ii}}\right\|_{1}C_{m-1}\left(\Sigma_{c}^{(i)}\right)\left(1 + \left\|\frac{\Sigma_{-i,i}}{\Sigma_{ii}}\right\|_{2}\right), \nonumber \\
    M^{*}_{i} &= e^{-\frac{1.5\ln\left(2\right)}{M_{i}\Sigma_{ii}}}, \nonumber \\
    \Sigma_{c}^{(i)} &= \Sigma_{-i,-i}-\frac{\Sigma_{-i,i}\Sigma_{i,-i}}{\Sigma_{ii}}.
\end{align}
\end{lemma}
\startproof{of Lemma}{lemma:mvn_cdf_grad_upperbound}
\begin{proofof}

For brevity, suppose that $\mu = 0$. By using the expressions (\ref{eq:7}), (\ref{eq:8}), and the formula (6) of \citep{Sungur1991}, we get:
\begin{align}
    \frac{\partial }{\partial x_{i}}\ln F(x) = \frac{f\left(x_{i};\Sigma_{ii}\right)F\left(x_{-i}-x_{i}\frac{\Sigma_{-i,i}}{\Sigma_{ii}};\Sigma_{c}^{(i)}\right)}{F\left(x;\Sigma\right)},
\end{align}
where:
\begin{align}
    \Sigma_{c}^{(i)} = \Sigma_{-i,-i}-\frac{\Sigma_{-i,i}\Sigma_{i,-i}}{\Sigma_{ii}}.
\end{align}

Suppose that there is $j\ne i$ such that $\Sigma_{ji}\ne 0$. Otherwise, the proof is trivial. However, without excluding this case, we may face a division by zero issue\footnote{This is because $M_{i}$ appears in the denominator of $M_{i}^{*}$, and $M_{i}=0$ when $\Sigma_{c}^{(i)}=\Sigma_{-i, -i}$.}.

We proceed with a proof by induction. In the univariate case $m=1$, for $x\geq 0$, we get:
\begin{align}
    \nabla_{x}\ln F(x) &\leq \frac{f(0)}{F(0)} \leq \frac{1}{\sqrt{\Sigma}} \leq \frac{1}{\sqrt{\Sigma}}\left(1 + \frac{|x|}{\sqrt{\Sigma}}\right).
\end{align}

For $x < 0$, we have:
\begin{align}
    \nabla_{x}\ln F(x) &= \frac{1}{\sqrt{\Sigma}}\frac{\phi(x/\sqrt{\Sigma})}{\Phi(x/\sqrt{\Sigma})} = \frac{1}{\sqrt{\Sigma}}/\frac{1 - \Phi(-x/\sqrt{\Sigma})}{\phi(-x/\sqrt{\Sigma})} \leq \nonumber \\ &\leq \underbrace{\frac{1}{\sqrt{\Sigma}}/\frac{2}{\sqrt{\left(-x/\sqrt{\Sigma}\right)^2+4}-x/\sqrt{\Sigma}}}_{\text{Theorem 3.1 of \citep{From2020}}} = \nonumber \\ 
    &= \frac{1}{\sqrt{\Sigma}}\frac{\sqrt{x^2/\Sigma+4}-x/\sqrt{\Sigma}}{2}\leq \frac{1}{\sqrt{\Sigma}}\left(1 + \frac{|x|}{\sqrt{\Sigma}}\right).
\end{align}

These results imply:
\begin{align}
    \nabla_{x}\ln F(x) &\leq \frac{1}{\sqrt{\Sigma}}\left(1 + \frac{|x|}{\sqrt{\Sigma}}\right)\leq  \nonumber \\ 
    &\leq \max\left(\frac{1}{\Sigma}, \frac{1}{\sqrt{\Sigma}}\right)\left(1 + |x|\right) = \nonumber \\
    &= C_{1}\left(\Sigma\right)\left(1 + \left\|x\right\|_{2}\right).
\end{align}

Suppose that the inequality holds for $m-1$, where $m>1$. Consider the following function:
\begin{align}
    h(t) = \mathbb{P}\left(X_{(-i)}\leq x_{(-i)}\mid X_{i} = t\right) = F\left(x_{-i}-t\frac{\Sigma_{-i,i}}{\Sigma_{ii}};\Sigma_{c}^{(i)}\right).
\end{align}

Notice that for $t\in\left[x_{i} - 1, x_{i}\right]$, we have:
\begin{align}\label{eq:lnht_upperbound}
    \left|\frac{\partial }{\partial t} \ln h(t) \right| &= \left|\sum\limits_{j\ne i}\frac{-\Sigma_{ji}}{\Sigma_{ii}}\frac{\partial }{\partial \left(x_{j}-t\frac{\Sigma_{ji}}{\Sigma_{ii}}\right)} \ln F\left(x_{-i}-t\frac{\Sigma_{-i,i}}{\Sigma_{ii}};\Sigma_{c}^{(i)}\right)\right| \leq \nonumber \\
    &\leq \underbrace{\left\|\frac{\Sigma_{-i,i}}{\Sigma_{ii}}\right\|_{1}C_{m-1}\left(\Sigma_{c}^{(i)}\right)\left(1 + \left\|x_{-i}-t\frac{\Sigma_{-i,i}}{\Sigma_{ii}}\right\|_{2}\right)}_{\text{Induction hypothesis}} \leq \nonumber \\
    &\leq \underbrace{\left\|\frac{\Sigma_{-i,i}}{\Sigma_{ii}}\right\|_{1}C_{m-1}\left(\Sigma_{c}^{(i)}\right)\left(1 + \left\|x_{-i}\right\|_{2}+\left\|\frac{\Sigma_{-i,i}}{\Sigma_{ii}}\right\|_{2}\times \left|t\right|\right)}_{\text{Triangle inequality}} \leq \nonumber \\ 
    &\leq \underbrace{\left\|\frac{\Sigma_{-i,i}}{\Sigma_{ii}}\right\|_{1}C_{m-1}\left(\Sigma_{c}^{(i)}\right)\left(1 + \left\|x\right\|_{2}+\left\|\frac{\Sigma_{-i,i}}{\Sigma_{ii}}\right\|_{2}\times \left(\left|x_{i}\right| + 1\right)\right)}_{\text{Inequalities }t\leq |x_{i}| + 1\text{ and }\left\|x_{-i}\right\|_{2}\leq \left\|x\right\|_{2}} \leq \nonumber \\
    &\leq  \underbrace{\left\|\frac{\Sigma_{-i,i}}{\Sigma_{ii}}\right\|_{1}C_{m-1}\left(\Sigma_{c}^{(i)}\right)\left(1 + \left\|\frac{\Sigma_{-i,i}}{\Sigma_{ii}}\right\|_{2}\right) \left(\left\|x\right\|_{2} + 1\right)}_{\text{Since }|x_{i}|=\left\|x_{i}\right\|_{2}\leq \left\|x\right\|_{2}} = \nonumber \\
    &= M_{i}\left(\left\|x\right\|_{2} + 1\right),
\end{align}
where:
\begin{align}
    M_{i} &= \left\|\frac{\Sigma_{-i,i}}{\Sigma_{ii}}\right\|_{1}C_{m-1}\left(\Sigma_{c}^{(i)}\right)\left(1 + \left\|\frac{\Sigma_{-i,i}}{\Sigma_{ii}}\right\|_{2}\right).
\end{align}

Consider $t\in \left[x_{i}-\delta_{i}, x_{i}\right]\subseteq \left[x_{i} - 1, x_{i}\right]$, where:
\begin{align}
    \delta_{i} &= \min\left(1, \frac{\ln\left(2\right)}{M_{i}\left(\left\|x\right\|_{2} + 1\right)}\right).
\end{align}

Hence, we get:
\begin{align}
    \left|\ln h(x_{i}) -  \ln h(t)\right| &\leq \underbrace{\left(x_{i} - t\right)}_{\text{Since }x_{i} \geq t} \times \underbrace{\max_{z\in \left[t, x_{i}\right]}\left|\frac{\partial }{\partial z} \ln h(z) \right|}_{\text{Largest derivative}} \leq \nonumber \\
    &\leq \delta_{i} \times \underbrace{M_{i}\left(\left\|x\right\|_{2} + 1\right)}_{\text{Formula (\ref{eq:lnht_upperbound})}} \leq \nonumber \\
    &\leq \underbrace{\ln\left(2\right)}_{\text{Since }\delta_{i}\leq \frac{\ln\left(2\right)}{M_{i}\left(\left\|x\right\|_{2} + 1\right)}}.
\end{align}

The last inequality implies:
\begin{align}\label{eq:ht_upperbound}
   \ln h(x_{i}) - \ln h(t) \le \ln(2) \implies h(t) \geq 0.5 h\left(x_{i}\right).
\end{align}

Further, notice that for $t\in \left[x_{i} - \delta_{i}, x_{i}\right]$, we have:
\begin{align}
    \frac{f(t;\Sigma_{ii})}{f(x_{i};\Sigma_{ii})} = e^{\frac{x_{i}^2 - t^2}{2\Sigma_{ii}}} = e^{\frac{\left(x_{i} - t\right)\times \left(x_{i} + t\right)}{2\Sigma_{ii}}} \geq e^{-\frac{\delta_{i}\times \left(2|x_{i}| + \delta_{i}\right)}{2\Sigma_{ii}}} \geq \nonumber \\
    \geq \underbrace{e^{-\frac{2\delta_{i}\left\|x\right\|_{2} + \frac{\ln\left(2\right)}{M_{i}}}{2\Sigma_{ii}}}}_{\text{Since }\delta_{i}\leq \frac{\ln\left(2\right)}{M_{i}}} \geq \underbrace{e^{-\frac{2\frac{\ln\left(2\right)}{M_{i}\left(\left\|x\right\|_{2} + 1\right)}\left\|x\right\|_{2} + \frac{\ln\left(2\right)}{M_{i}}}{2\Sigma_{ii}}}}_{\text{Since }\delta_{i}\leq \frac{\ln\left(2\right)}{M_{i}\left(\left\|x\right\|_{2} + 1\right)}} \geq e^{-\frac{\frac{3\ln\left(2\right)}{M_{i}}}{2\Sigma_{ii}}} \geq e^{-\frac{1.5\ln\left(2\right)}{M_{i}\Sigma_{ii}}}.
\end{align}

This result implies:
\begin{align}\label{eq:ft_upperbound}
    f(t;\Sigma_{ii})\geq f(x_{i};\Sigma_{ii}) M^{*}_{i},
\end{align}
where:
\begin{align}
    M^{*}_{i} = e^{-\frac{1.5\ln\left(2\right)}{M_{i}\Sigma_{ii}}}.
\end{align}

By using the formulas (\ref{eq:ht_upperbound}) and (\ref{eq:ft_upperbound}), we get:
\begin{align}
    F(x;\Sigma) &= \underbrace{\int_{-\infty}^{x_{i}} f(t;\Sigma_{ii})\mathbb{P}\left(X_{(-i)}\leq x_{(-i)}\mid X_{i} = t\right) dt}_{\text{Law of total probability}} = \nonumber \\
    &= \int_{-\infty}^{x_{i}} f(t;\Sigma_{ii})h(t) dt \geq \nonumber \\ 
    &\geq \int_{x_{i}-\delta_{i}}^{x_{i}} f(t;\Sigma_{ii})h(t) dt \geq \nonumber \\ 
    &\geq \underbrace{\int_{x_{i}-\delta_{i}}^{x_{i}} f(x_{i};\Sigma_{ii}) M^{*}_{i}\times 0.5h(x_{i}) dt}_{\text{Formulas (\ref{eq:ht_upperbound}) and (\ref{eq:ft_upperbound})}} = \nonumber \\
    &= \delta_{i}f(x_{i};\Sigma_{ii}) M^{*}_{i}\times 0.5h(x_{i}) = \nonumber \\ 
    & = \left(0.5\delta_{i}M_{i}^{*}\right)\times f(x_{i};\Sigma_{ii})F\left(x_{-i}-x_{i}\frac{\Sigma_{-i,i}}{\Sigma_{ii}};\Sigma_{c}^{(i)}\right).
\end{align}

This inequality implies:
\begin{align}
    \frac{\partial }{\partial x_{i}} \ln F(x) &\leq \frac{f\left(x_{i};\Sigma_{ii}\right)F\left(x_{-i}-x_{i}\frac{\Sigma_{-i,i}}{\Sigma_{ii}};\Sigma_{c}^{(i)}\right)}{\left(0.5\delta_{i}M_{i}^{*}\right)\times f(x_{i};\Sigma_{ii})F\left(x_{-i}-x_{i}\frac{\Sigma_{-i,i}}{\Sigma_{ii}};\Sigma_{c}^{(i)}\right)} = \nonumber \\
    &= \frac{1}{0.5\delta_{i}M_{i}^{*}}\leq c_{i}\left(1 + \left\|x\right\|_{2}\right),
\end{align}
where:
\begin{align}
    c_{i} = \frac{2}{M_{i}^{*}}\left(1+\frac{M_{i}}{\ln\left(2\right)}\right).
\end{align}

Therefore, we finally get:
\begin{align}
    \left\|\nabla_{x} \ln F(x)\right\|_{2} &\leq \sum\limits_{i=1}^{m}\frac{\partial }{\partial x_{i}} \ln F(x) \leq C_{m}\left(\Sigma\right)\left(1 + \left\|x\right\|_{2}\right),
\end{align}
where:
\begin{align}
    C_{m}\left(\Sigma\right) = \sum\limits_{i=1}^{m} c_{i}.
\end{align}

\end{proofof}

% Ограничение сверху производных вероятностей
\begin{lemma}\label{lemma:mvn_cdf_hessian_upperbound}
The following inequalities hold:
\begin{align}
    \left|\frac{\partial^2 }{\partial x_{i} \partial x_{j}} \ln F(x)\right| &\leq C_{m}\left(\Sigma\right)\left(1 + \left\|x - \mu\right\|_{2}\right)^2, \\
    \left|\frac{\partial}{\partial \Sigma_{ij}} \ln F(x)\right| &\leq C_{m}\left(\Sigma\right)\left(1 + \left\|x - \mu\right\|_{2}\right)^2,
\end{align}
where $C_{m}\left(\Sigma\right)$ is a function of $\Sigma$ whose expression is omitted for brevity.
\end{lemma}
\startproof{of Lemma}{lemma:mvn_cdf_hessian_upperbound}
\begin{proofof}
The proof proceeds by the same steps as the proof of Lemma \ref{lemma:mvn_cdf_grad_upperbound} and is therefore omitted for brevity.
\end{proofof}

\section{Conclusion}\label{sec:conclusion}

We have proposed the \textbf{mnorm} package which allows one to calculate and differentiate the conditional MVN densities and probabilities. The ability to calculate the gradients with respect to various parameters of MVN densities and probabilities is the main contribution of the package. Simulated data analysis suggests that analytical gradients notably outperform numeric gradients in terms of speed. The difference is especially great for probabilities when $m_{d}\leq 5$ because the derivatives are calculated via fast bivariate and trivariate computation methods. Moreover, the analytical derivatives with respect to the parameters associated with the given (conditioned) parts of the conditional distribution are very fast, since they are just functions of the derivatives with respect to other parameters. Therefore, we hope that our package will substantially ease the implementation of maximum-likelihood estimators involving calculation of MVN densities and probabilities. The package may be especially useful for econometric models involving the conditional MVN probabilities, for example, multivariate and multinomial sample selection and endogenous switching models \citep{{abort}, {Dolgikh2023A}, {Dolgikh2023B}, {Dolgikh2024}, {Dolgikh2026gender}, {Dolgikh2026}}.

Also, according to the results of simulated data analysis, \code{pmnorm} notably outperforms other popular alternatives in terms of calculation speed of the bivariate and trivariate MVN probabilities. Moreover, \code{dmnorm} is slightly faster than the alternative functions for calculation of the MVN densities.

Finally, we have provided convenient bounds on the derivatives of the log-probabilities of the MVN distribution.

% Literature
\bibliographystyle{plainnat}
\bibliography{literature}

\end{document}